\documentclass{llncs}

\usepackage{epsfig}
\usepackage{psfig}
\usepackage{anysize}

\newcommand{\cP}{\mbox{${\cal P}$}}

\long\def\comment#1{}

\begin{document}

\title{High-Throughput SNP Genotyping by SBE/SBH\thanks{Work 
supported in part by a Faculty Large Research Grant from the 
University of Connecticut Research Foundation.}}

\author{
Ion I. M\u{a}ndoiu
\and
Claudia Pr\u{a}jescu
}

\institute{CSE Department, University of Connecticut\\
371 Fairfield Rd., Unit 2155, Storrs, CT 06269-2155\\
\email{\{ion.mandoiu,claudia.prajescu\}@uconn.edu} 
}

\maketitle


\begin{abstract}
Despite much progress over the past decade, 
current Single Nucleotide Polymorphism (SNP) genotyping
technologies still offer an insufficient degree of multiplexing 
when required to handle user-selected sets of SNPs. 
In this paper we propose a new genotyping assay architecture combining
multiplexed solution-phase single-base extension (SBE) reactions 
with sequencing by hybridization (SBH) using universal DNA arrays 
such as all $k$-mer arrays.
In addition to PCR amplification of genomic DNA, SNP genotyping 
using SBE/SBH assays involves the following steps:
(1) Synthesizing primers complementing
the genomic sequence immediately preceding SNPs of interest; 
(2) Hybridizing these primers with the genomic DNA;
(3) Extending each primer by a single base using 
polymerase enzyme and dideoxynucleotides labeled
with 4 different fluorescent dyes; and finally
(4) Hybridizing extended primers to a universal DNA array 
and determining the identity of the bases that extend 
each primer by hybridization pattern analysis.
Under the assumption of perfect hybridization, unambiguous 
genotyping of a set of SNPs requires selecting 
primers upstream of the SNPs such that each 
primer hybridizes to at least one array probe that 
hybridizes to no other primer that can be extended by a common base.
Our contributions include a study of multiplexing algorithms 
for SBE/SBH genotyping assays and preliminary experimental results
showing the achievable tradeoffs between the number of 
array probes and primer length on one hand and the 
number of SNPs that can be assayed simultaneously on the other. 
We prove that the problem of selecting a maximum size subset of SNPs 
that can be unambiguously genotyped in a single SBE/SBH assay 
is NP-hard, and propose efficient heuristics 
with good practical performance.  
Our heuristics take into account the freedom 
of selecting primers from both strands of the genomic DNA
as well as the presence of disjoint allele sets among genotyped SNPs. 
In addition, our heuristics can enforce user-specified 
redundancy constraints facilitating reliable genotyping 
in the presence of hybridization errors.
Simulation results on datasets both randomly generated and 
extracted from the NCBI dbSNP database 
suggest that the SBE/SBH architecture provides 
a flexible and cost-effective alternative to 
genotyping assays currently used in the industry, 
enabling genotyping of up to hundreds 
of thousands of user-specified SNPs per assay.
\end{abstract}

\section{Introduction}
\label{sec.intro}

After the completion of the Human Genome Project has provided
a blueprint of the DNA present in each human cell \cite{HGP01,HGP04},
genomics research is now focusing on the study of DNA variations
that occur between individuals, seeking to understand
how these variations confer susceptibility to 
common diseases such as diabetes or cancer.
The most common form of genomic variation
are the so called {\em single nucleotide polymorphisms} (SNPs),
i.e., the presence of different DNA nucleotides, 
or {\em alleles}, at certain chromosomal locations.
The vast majority of SNPs are {\em bi-allelic}, i.e., only two of 
the four possible DNA bases are observed at the SNP locus. 
Since human cells contain two copies of each chromosome (with the
exception of sex chromosomes in males), both SNP alleles may be 
present in the DNA of an individual. Determining the identity of 
alleles present in a DNA sample at a given set of SNP loci is called 
{\em SNP genotyping}.

The continuous progress in high-throughput genomic
technologies has resulted in numerous SNP genotyping 
platforms combining a variety of
allele discrimination techniques
(sequencing, direct hybridization, primer extension,
allele-specific PCR, ligation, and cleavage, etc.),
detection mechanisms
(fluorescence, mass spectrometry, etc.)
and reaction formats (solution phase, solid support,
bead arrays), see, e.g., \cite{Jenkins02,Kwok01} for 
comprehensive reviews.
However, current technologies still offer
an insufficient degree of multiplexing (below
10,000 SNPs per assay) for 
fully-powered genome wide disease 
association studies that require genotyping of 
large sets of user-selected SNPs \cite{Carlson04}.
The highest throughput is currently achieved by high-density 
mapping arrays produced by Affymetrix, which can simultaneously 
genotype a fixed set of about 250,000 {\em manufacturer selected} SNPs 
per array.
Genotyping a comparable number of user-specified set of SNPs 
would require an expensive and time-consuming re-design of 
array probes as well as a difficult re-engineering of the 
primer-ligation amplification protocol.

Among technologies that allow genotyping of custom sets of SNPs one of 
the most successful ones is the use of DNA tag arrays 
\cite{Brenner97,Gerry99,Hirschhorn2000,affy_patent}.
DNA tag arrays consist of a set of DNA strings called {\em tags}, 
designed such that each tag hybridizes strongly to 
its own {\em antitag} (Watson-Crick complement),
but to no other antitag.
The flexibility of tag arrays comes from combining
solid-phase hybridization with the high sensitivity of
single-base extension reactions, which
has also been used for SNP genotyping in combination with
MALDI-TOF mass spectrometry \cite{Aumann03}.
A typical assay based on tag arrays performs
SNP genotyping using the following
steps \cite{BenDor03,Hirschhorn2000}:
(1) A set of {\em reporter probes}
is synthesized by ligating antitags to the $5'$ end of primers complementing
the genomic sequence immediately preceding the SNPs of interest.
(2) Reporter probes are hybridized in solution with the genomic sample.
(3) The hybridized $3'$ (primer) end of reporter probes is extended 
by a single base in a reaction using 
the polymerase enzyme and dideoxynucleotides fluorescently labeled
with 4 different dyes.
(4) Reporter probes are separated from the template DNA
and hybridized to a tag array.
(5) Finally, fluorescence levels   
are used to determine the identity of the extending dideoxynucleotides.
Commercially available tag arrays have between 2,000 and 10,000 tags 
\cite{genflex,truetag}. The number of SNPs that can be 
genotyped per array is typically smaller than the number of tags 
since some of the tags must remain unassigned due to cross-hybridization with the primers 
\cite{BenDor03,lncs_tcsb05}. 
Another factor limiting the wider use of tag arrays 
is the relatively high cost of synthesizing the reporter probes, which have a 
typical length of 40 nucleotides. 

In the $k$-mer array format \cite{Dramanac87},
all $4^k$ DNA probes of length $k$ are spotted or
synthesized on the solid array substrate
(values of $k$ of up to $10$ are feasible with
current high-density in-situ synthesis technologies).
This format was originally proposed for performing
{\em sequencing by hybridization (SBH)}, 
which seeks to reconstruct an unknown DNA sequence based on
its $k$-mer spectrum \cite{Pevzner00}. However, the sequence
length for which unambiguous reconstruction is possible
with high probability is surprisingly small 
\cite{Pevzner94}, and, despite
several suggestions for improvement, such as 
the use of gapped probes \cite{Heath01} and 
pooling of target sequences \cite{Hubbell01},
the SBH scheme has not become practical so far.

\begin{figure}[t]
\begin{center}
    \begin{minipage}[b]{0.4\linewidth}
      \centering \psfig{file=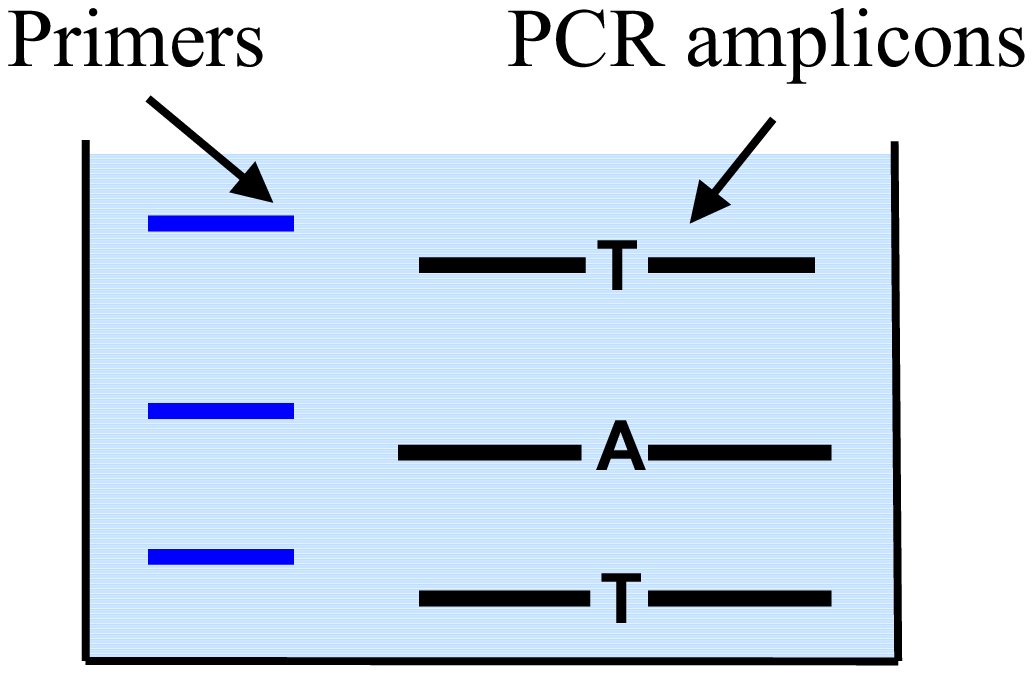,width=2.0in}
      \centerline{(a)}
    \end{minipage}
    \hspace{0.3cm}
    \begin{minipage}[b]{0.4\linewidth}
      \centering \psfig{file=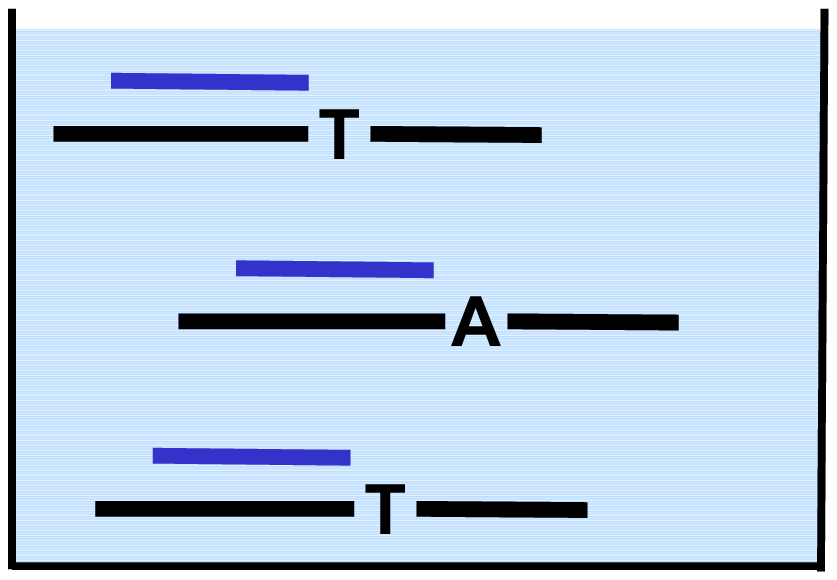,width=2.0in}
      \centerline{(b)}
    \end{minipage}\\[5mm]
    \begin{minipage}[b]{0.4\linewidth}
      \centering \psfig{file=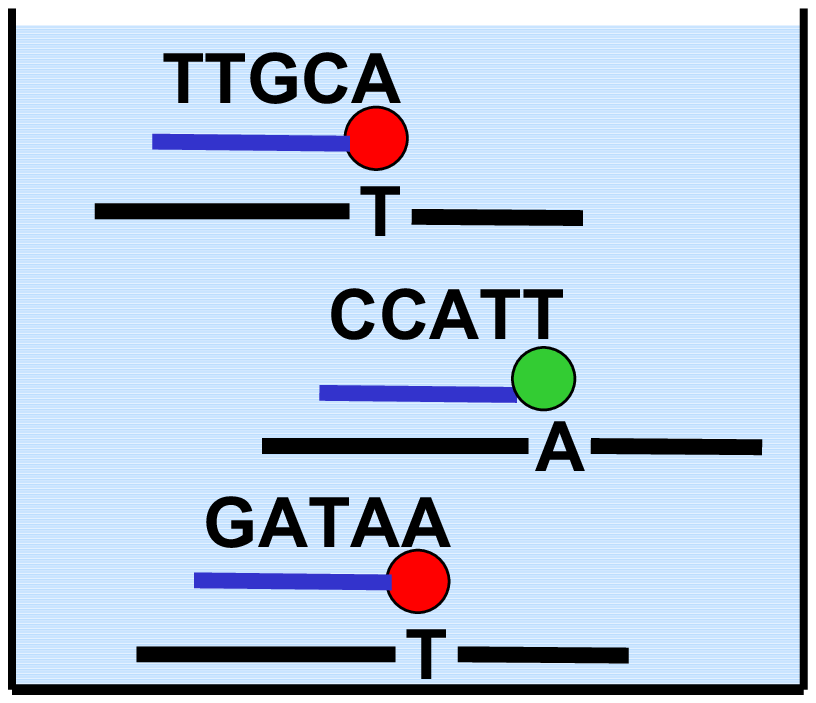,width=2.0in}
      \centerline{(c)}
    \end{minipage}
    \hspace{0.3cm}
    \begin{minipage}[b]{0.4\linewidth}
      \centering \psfig{file=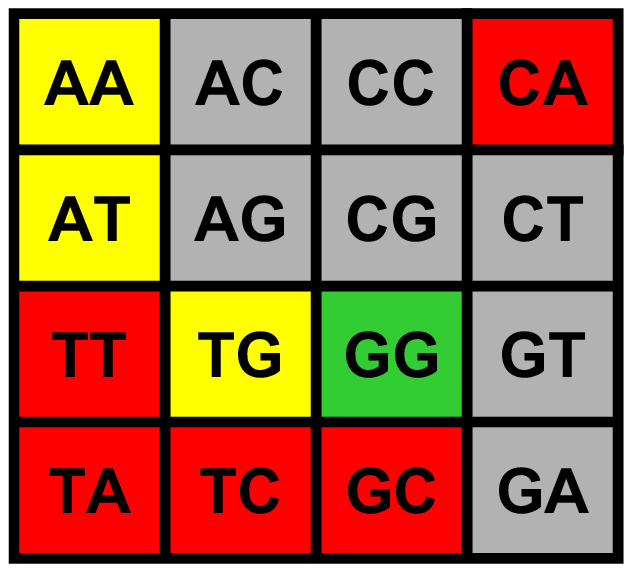,width=2.0in}
      \centerline{(d)}
    \end{minipage}

\caption{\label{k-mer}\sl SBE/SBH assay:
(a) Primers complementing genomic sequence upstream of each SNP locus are mixed 
in solution with the genomic DNA sample.
(b) Temperature is lowered allowing primers to hybridize to the genomic DNA.
(c) Polymerase enzyme and dideoxynucleotides labeled
with 4 different fluorescent dyes are added to the solution, causing 
each primer to be extended by a nucleotide complementing the SNP allele.
(d) Extended primers are hybridized to a universal DNA array (an all $k$-mer array for
$k$=2 is shown)
and SNP genotypes are determined by analyzing the resulting hybridization pattern.
Under the assumption of perfect hybridization, unambiguous
genotyping of the SNPs requires that each
primer hybridizes to at least one array probe that
hybridizes to no other primer that can be extended by a common base.}
\end{center}
\end{figure}

In this paper we propose a new genotyping assay architecture combining
multiplexed solution-phase single-base extension (SBE) reactions      
with sequencing by hybridization (SBH) using universal DNA arrays    
such as all $k$-mer arrays.   
SNP genotyping using SBE/SBH assays requires
the following steps (see Figure \ref{k-mer}):
(1) Synthesizing primers complementing
the genomic sequence immediately preceding SNPs of interest; 
(2) Hybridizing primers with the genomic DNA;
(3) Extending each primer by a single base using 
polymerase enzyme and dideoxynucleotides labeled
with 4 different fluorescent dyes; and finally
(4) Hybridizing extended primers to a universal DNA array 
and determining the identity of the bases that extend 
each primer by hybridization pattern analysis.

To the best of our knowledge the combination of the two technologies 
in the context of SNP genotyping has not been explored thus far.    
The most closely related genotyping assay
is the generic Polymerase Extension Assay (PEA) recently 
proposed in \cite{Sharan05}. In PEA, 
short amplicons containing the SNPs of interest are hybridized to 
an all $k$-mers array of {\em primers} that are subsequently 
extended via single-base extension reactions. 
Hence, in PEA the SBE reactions take place on solid support, similar 
to {\em arrayed primer extension} (APEX) assays which use 
SNP specific primers spotted on the array \cite{Tonisson00}.

As in \cite{Hubbell01}, the SBE/SBH assay leads to
high array probe utilization since we hybridize to the array
a large number of short extended primers.  However,
the main power of the method lies in the fact that
the sequences of the labeled oligonucleotides hybridized to the array
are a priori known (up to the identity of extending nucleotides).
While genotyping with SBE/SBH assays uses similar
general principles as the PEA assays proposed
in \cite{Sharan05}, there are also significant differences.
A major advantage of SBE/SBH 
is the much shorter length of extended primers
compared to that of PCR amplicons used in PEA. 
A second advantage is that {\em all} probes hybridizing to 
an extended primer are informative in SBE/SBH assays, regardless 
of array probe length (in contrast, only probes hybridizing with a substring 
containing the SNP site are informative in PEA assays). 
As shown by the experimental results in Section \ref{sec.results}
these advantages translate into an increase by 
orders of magnitude in multiplexing rate compared to the 
results reported in \cite{Sharan05}.
We further note that PEA's effectiveness crucially depends on 
the ability to amplify very short (preferably 40bp or less) 
genomic fragments spanning the SNP loci of interest. 
This limits the achievable degree of multiplexing in 
PCR amplification \cite{apbc05}, making PCR amplification 
the main bottleneck for PEA assays. Full flexibility in 
picking PCR primers is preserved in SBE/SBH assays.

The rest of the paper is organized as follows.  
In Section \ref{sec.formulations} we formalize two problems 
that arise in genotyping large sets of SNPs using SBE/SBH assays: 
the problem of partitioning a set of SNPs into the minimum number of 
``decodable'' subsets, i.e., subsets of SNPs that can
be unambiguously genotyped using a single SBE/SBH assay,
and that of finding a maximum decodable subset of a 
given set of SNPs.  We also establish hardness results for 
the latter problem.
In Section \ref{sec.algos} we propose several efficient heuristics. 
Finally, in Section \ref{sec.results} we present experimental results 
on both randomly generated datasets and instances 
extracted from the NCBI dbSNP database, 
exploring achievable tradeoffs 
between the type/number of array probes and primer length on 
one hand and number of SNPs that can be assayed per array on the other.
Our results suggest that the SBE/SBH architecture provides     
a flexible and cost-effective alternative to
genotyping assays currently used in the industry,
enabling genotyping of up to hundreds           
of thousands of user-selected SNPs per assay.

\section{Problem Formulations and Complexity} 
\label{sec.formulations}

A set of SNP loci can be unambiguously genotyped by SBE/SBH 
if every combination of SNP genotypes yields a different 
hybridization pattern (defined as the vector of 
dye colors observed at each array probe). 
To formalize the requirements of unambiguous genotyping, 
let us first consider a simplified SBE/SBH assay 
consisting of four parallel {\em single-color} SBE/SBH 
reactions, one for each possible SNP allele.  
Under this scenario, 
only one type of dideoxynucleotide is added to each SBE reaction,
corresponding to the complement of the tested SNP allele. 
Therefore, a primer is extended in such a reaction if  
the tested allele is present at the SNP locus probed by the primer, 
and is left un-extended otherwise. 

Let $\cP$ be the set of primers used in a single-color SBE/SBH 
reaction involving dideoxynucleotide $e\in \{$A,C,G,T$\}$.
From the resulting hybridization pattern we must be able to infer 
for every $p\in \cP$ whether or not $p$ was extended by $e$.  
The extension of $p$ by $e$ will result in a fluorescent signal 
at all array probes that hybridize with $pe$.  
However, some of these probes can give a fluorescent signal 
even when $p$ is not extended  by $e$, 
due to hybridization to other extended primers.  
Since in the worst case {\em all} other primers are 
extended, it must be the case that at least one of the probes 
that hybridize to $pe$ does not hybridize to any other 
extended primer.

Formally, let $X\subset \{A,C,G,T\}^*$ be the set of array probes. 
For every string $y\in \{A,C,G,T\}^*$, let the 
{\em spectrum of $y$ in $X$}, denoted $Spec_X(y)$, be the 
set of probes of $X$ that hybridize with $y$.  
Under the assumption of perfect hybridization, $Spec_X(y)$ 
consists of those probes of $X$ that are Watson-Crick 
complements of substrings of $y$. 
Then, a set of primers $\cP$ is said to be {\em decodable} with 
respect to extension $e$ if and only if, for every $p\in \cP$,
\begin{equation}\label{1-color-weak}
Spec_X(pe) \setminus \bigcup_{p'\in \cP\setminus\{p\}}Spec_X(p'e) \neq \emptyset 
\end{equation}

Decoding constraints (\ref{1-color-weak}) can be directly extended 
to 4-color SBE/SBH experiments, in which each type of extending base 
is labeled by a different fluorescent dye.  As before, let $\cP$ be the 
set of primers, and, for each primer $p\in \cP$, let $E_p\subseteq \{A,C,G,T\}$ 
be the set of possible extensions of $p$, i.e., Watson-Crick complements 
of corresponding SNP alleles. 
If we assume that any combination of dyes can be detected at an 
array probe location, unambiguous decoding is guaranteed  
if, for every $p\in \cP$ and every extending nucleotide $e\in E_p$,
\begin{equation}\label{4-color-weak}
Spec_X(pe) \setminus \bigcup_{p'\in \cP\setminus\{p\}, e\in E_{p'}}Spec_X(p'e) 
\neq \emptyset
\end{equation}

In the following, we refine (\ref{4-color-weak}) to improve 
practical reliability of SBE/SBH assays.  More precisely, 
we impose additional constraints on the set of probes considered 
to be {\em informative} for each SNP allele. 
First, to enable reliable genotyping of genomic 
samples that contain SNP alleles at very different concentrations 
(as a result of uneven efficiency in the PCR amplification step or 
of pooling DNA from different individuals), we require that 
a probe that is informative for a certain SNP locus must not hybridize 
to primers corresponding to different SNP loci, 
{\em regardless of their extension}.
Second, since recent studies by Naef et al. \cite{Naef03} suggest 
that fluorescent dyes can significantly interfere with oligonucleotide 
hybridization on solid support, possibly destabilizing hybridization to 
a complementary probe on the array, in this paper we 
use a conservative approach and require that each 
probe that is informative for a certain SNP allele 
must hybridize to a strict substring of the corresponding primer.
On the other hand, informative probes are still required 
not to hybridize with any other extended primer, even if such  
hybridizations involve fluorescently labeled nucleotides.
Finally, we introduce a {\em decoding redundancy} parameter $r\ge 1$, 
and require that each SNP have at least $r$ informative 
probes, i.e., probes that hybridize to the corresponding primer 
but do not hybridize to any other extended primer.  
Such a redundancy constraint facilitates 
reliable genotype calling in the presence of hybridization errors.
Clearly, the larger the value of $r$, the more hybridization 
errors that can be tolerated. If a simple majority voting 
scheme is used for making allele calls, the assay can 
tolerate up to $\lfloor r/2\rfloor$ hybridization errors 
involving the $r$ informative probes of each SNP. 
Furthermore, since the informative probes of a SNP 
are required to hybridize {\em exclusively} with the primer 
corresponding to the SNP, the redundancy requirement 
provides a powerful mechanism for detecting and 
gauging the extent of hybridization errors. Indeed, each 
unintended hybridization at an informative probe for a 
bi-allelic SNP has a dye complementary to one of the SNP alleles 
with probability of only 1/2, and the probability that $k$ such errors 
pass undetected decreases exponentially in $k$.

The refined set of constraints is captured by the following definition, where,
for every primer $p\in \{A,C,G,T\}^*$ and set of extensions 
$E\subseteq \{A,C,G,T\}$, we let 
\[
 Spec_X(p,E)=\bigcup_{e\in E} Spec_X(pe)
\]
\begin{definition}\label{def.strong-primer}
A set of primers $\cP$ is said to be {\em strongly $r$-decodable} with 
respect to extension sets $E_p$, $p\in \cP$, if and only if, 
for every $p\in \cP$, 
\begin{equation}\label{4-color-strong}
\left| Spec_X(p) \setminus 
 \bigcup_{p'\in \cP\setminus\{p\}} Spec_X(p',E_{p'}) \right| \geq r
\end{equation}
\end{definition}
Note that testing whether or not a given set of primers is strongly 
$r$-decodable can be easily accomplished in time linear in the total length 
of the primers. 

Genotyping a large set of SNPs will, in general, require more than 
one SBE/SBH assay. This rises the problem of partitioning 
a given set of SNPs into the smallest number of subsets that 
can each be genotyped using a single SBE/SBH assay. 
For each SNP locus there are typically two different primers 
that can be used for genotyping.  
As shown in \cite{lncs_tcsb05} for the case of SNP genotyping using tag 
arrays, exploiting this degree of freedom significantly increases 
achievable multiplexing rates.  Therefore, 
we next extend our definitions to capture this degree of freedom. 
Let $P_i$ be the {\em pool of primers} that can be used to genotype 
the SNP at locus $i$.  Similarly to Definition \ref{def.strong-primer}, 
we have:
\begin{definition}\label{def.strong-pool}
A set of primer pools $\cP=\{P_1,\ldots,P_n\}$ 
is said to be {\em strongly $r$-decodable} 
if and only if there is a primer $p_i$ in each pool $P_i$ 
such that $\{p_1,\ldots,p_n\}$ is 
strongly $r$-decodable with respect to the respective 
extension sets $E_{p_i}$, $i=1,\ldots,n$.
\end{definition}
Primers $p_1,p_2,\ldots,p_n$ above are called the {\em representative primers} of
pools $P_1,P_2,\ldots,P_n$, respectively. 
The SNP partitioning problem can then be formulated as follows:

\bigskip
\noindent
{\bf 
Minimum Pool Partitioning Problem (MPPP):} 
{\em Given primer pools $\cP=\{P_1,\ldots,P_n\}$, 
associated extension sets $E_p$, $p\in \cup_{i=1}^n P_i$, 
probe set $X$,
and redundancy $r$, 
find a partitioning of $\cP$ into the minimum number of 
strongly $r$-decodable subsets. 
}
\bigskip

A natural strategy for solving MPPP, 
similar to the well-known greedy algorithm for the set cover problem, 
is to find a maximum strongly $r$-decodable subset of pools, 
remove it from $\cP$, and then repeat 
the procedure until no more pools are left in $\cP$.  
This greedy strategy for solving MPPP has been shown to 
empirically outperform other algorithms for solving the similar 
partitioning problem for PEA assays \cite{Sharan05}. 
In the case of SBE/SBH, the optimization involved in the 
main step of the greedy strategy is formalized as follows:

\bigskip
\noindent
{\bf 
Maximum $r$-Decodable Pool Subset Problem (MDPSP):} 
{\em Given primer pools $\cP=\{P_1,\ldots,P_n\}$, 
associated extension sets $E_p$, $p\in \cup_{i=1}^n P_i$, 
probe set $X$,
and redundancy $r$, 
find a strongly $r$-decodable subset $\cP' \subseteq \cP$ of maximum size. 
In addition, for each pool
$P_i \in \cP'$, find its representative primer.
}
\bigskip

Unfortunately, as shown in next theorem, MDPSP is NP-hard 
even for the case when the redundancy parameter is 1 
and each pool has exactly one primer.

\begin{theorem}\label{theorem.MDPSPhard}
MDPSP is NP-hard, 
even when restricted to instances with $r=1$ and $|P|=1$ for every  $P\in\cP$.
\end{theorem}

\begin{proof}
We will use a reduction from the {\em maximum induced matching} 
problem in bipartite graphs, which is defined as follows:

\bigskip
\noindent
{\bf Maximum Induced Matching (MIM) Problem in Bipartite Graphs:}
{\em Given a bipartite graph $G=(U \cup V, E)$,
find maximum size subsets $U' \subseteq U$, $V' \subseteq V$, with 
$|U'|=|V'|$ such that the subgraph of $G$ 
induced by $U' \cup V'$ is a matching.
}
\bigskip

The MIM problem in bipartite graphs is known to be NP-hard 
even for graphs with maximum degree 3 \cite{Lozin02}.
Let $G=(U \cup V, E)$ be such a bipartite graph with maximum degree 3. 
Without loss of generality we may assume that every vertex 
in $G$ has degree at least 1.
We will denote by $N(u)$ the {\em neighborhood} of vertex $u \in U \cup V$, 
i.e., the set of vertices adjacent with $u$ in $G$.

We construct an instance of MDPSP as follows: 
Let $r=1$ and $l=\lceil \log_2 |V| \rceil$. 
For every $v \in V$ we add to $X$ a distinct probe $x_v \in \{$A,T$\}^l$;
note that this can be done since $|\{$A,T$\}^l|=2^l>|V|$ by our choice of $l$. 
For every $u \in U$, with neighborhood $N(u)=\{v_{1},v_{2},v_{3}\}$, we 
construct a primer $p_u=x_{v_1}$C$x_{v_2}$C$x_{v_3}$ and 
set $P_u=\{ p_u \}$. We use a similar construction for 
vertices $u\in U$ with only 1 or 2 neighbors.
Note that in each case the pool $P_u$ consists of a single primer $p_u$ of 
length at most $3l+2$.
For each constructed primer $p$, the set of possible 
extensions is defined as $E_{p}=\{$G,C$\}$. 
Since the probes of $X$ contain only A's and T's, for every primer $p_u$, 
$u\in U$, 
\[
  Spec_X(p_u,E_{p_u}) = Spec_X(p_u) = \{x_v \in X |\ v \in N(u)\}
\]

Let $U' \subseteq U$, $V' \subseteq V$, $|U'|=|V'|$, be subsets of 
vertices such that  $U' \cup V'$ induces a matching in $G$.
Let $\cP' = \{P_{u} |\ u \in U'\}$.
For every $u\in U'$, exactly one of $u$'s neighbors, denoted $v_{u}$, 
appears in $V'$, because $U'\cup V'$ induces a matching.
Furthermore, for each $u' \in U'\setminus \{u\}$, $(u',v_{u}) \notin E$, 
and therefore $x_{v_u}\notin Spec_X(p_{u'},E_{p_{u'}})$.
Thus, for every $u\in U'$,
\[
  x_{v_u}\in Spec_X(p_u) \setminus 
   \bigcup_{\{p_{u'}\} \in \cP'\setminus\{p_u\}}Spec_X(p_{u'},E_{p_{u'}}) 
\]
which means that $\cP'$ is a strongly 1-decodable subset of pools of the same size 
as the induced matching of $G$.

Conversely, let $\cP'$ be a strongly 1-decodable subset of $\cP$, 
and let $U' = \{u \in U |\ \{p_u\} \in \cP'\}$. 
Since $\cP'$ is 1-decodable, for every primer $p_u$ with $\{p_u\} \in \cP'$, 
there must exist a probe $x \in X$ such that 
$x \in Spec_X(p_u)$ and $x \notin Spec_X(p_{u'},E_{p_{u'}})$  
for every $\{p_{u'}\} \in \cP'\setminus \{p_u\}$.
Because $Spec_X(p_u)=\{x_v \in X |\ v \in N(u)\}$, 
it follows that every vertex $u \in U'$ has a neighbor $v \in V$ 
that is not a neighbor of any other $u' \in U'\setminus \{u\}$. 
Let $v_u$ be such a neighbor (pick $v_u$ arbitrarily if more than 
one vertex in $V$ satisfies above property), and 
let $V'=\{v_u |\ u\in U' \}$.
It is clear that $U'\cup V'$ induce a matching of 
size $|\cP'|$ in $G$.

Thus, for every integer $k$, there is a one-to-one correspondence 
between induced matchings of size $k$ in $G$
and strongly 1-decodable subsets of $k$ pools in the constructed 
instance of MDPSP, and NP-hardness of MDPSP follows.
\end{proof}

The reduction in the proof of Theorem \ref{theorem.MDPSPhard} preserves the size 
of the optimal solution, and therefore
any hardness of approximation result for the MIM in bipartite graphs 
will also hold for MDPSP, 
even when restricted to instances with $r=1$ and $|P|=1$ for every  $P\in\cP$.
Since Duckworth et al. \cite{Zito05} proved that it is NP-hard to 
approximate MIM in bipartite graphs with maximum degree 3
within a factor of 6600/6659, 
we get:

\begin{theorem}\label{corol.APX-hard}
It is NP-hard to approximate 
MDPSP within a factor of 6600/6659, 
even when restricted to instances with $r=1$ and $|P|=1$ for every  $P\in\cP$.
\end{theorem}

\section{Algorithms} 
\label{sec.algos}

In this section we describe three heuristic approaches to MDPSP.
The first one is a naive greedy algorithm that sequentially evaluates 
the primers in the given pools in an arbitrary order. 
The algorithm picks a primer $p$ to be the representative of pool $P\in\cP$ 
if $p$ together with the representatives already picked satisfy condition 
(\ref{4-color-strong}). The pseudocode of this algorithm, which we 
refer to as Sequential Greedy, is given in Figure \ref{fig.seq}.

\begin{figure}[t]             
{\footnotesize
\fbox{
\begin{minipage}{\textwidth}             
\begin{tabbing}
\hspace*{5mm}\=\hspace{5mm}\=\hspace{5mm}\=\hspace{5mm}\=\hspace{5mm}\=\hspace{5mm}\=  \kill
{\tt Input:} Pools $\cP=\{P_1,\dots,P_n\}$, 
extension sets $E_p$, $p\in \cup_{i=1}^n P_i$, probe set $X$, and redundancy $r$\\
{\tt Output:} Strongly $r$-decodable subset of pools $\cP' \subseteq \cP$ 
and set $R$ of representative primers for the pools in \cP' \\
\rule[3pt]{1.0\textwidth}{0.3pt}\\
0. $\cP' \leftarrow \emptyset$,\ \ $R \leftarrow \emptyset$ \\
1. {\tt For each} $P \in \cP$ {\tt do}\\
2. \>\> {\tt For each} $p \in P$ {\tt do}\\
3. \>\>\>  {\tt If} $R \cup \{p\}$ satisfies (\ref{4-color-strong})  \\
   \>\>\>  {\tt Then} \\ 
4. \>\>\>\>  $\cP' \leftarrow \cP' \cup P$ \\ 
4. \>\>\>\>  $R \leftarrow R \cup \{p\}$ \\ 
5. \>\>\>\>  Exit inner {\tt For}  \\ 
   \>\>\>  {\tt End If} \\ 
   \>\> {\tt End For} \\ 
   {\tt End For} \\
\end{tabbing}
\end{minipage}}
}
\caption{\label{fig.seq} The Sequential Greedy algorithm.}
\end{figure}

The next two algorithms are inspired by the Min-Greedy algorithm in 
\cite{Zito05}, which approximates MIM in $d$-regular graphs 
within a factor of $d-1$. For the MIM problem, the Min-Greedy algorithm 
picks at each step a vertex $u$ of minimum 
degree and a vertex $v$, which is a minimum degree 
neighbor of $u$. All the neighbors of $u$ and $v$ are deleted and 
the edge $(u,v)$ is added to the induced matching. 
The algorithm stops when the graph becomes empty.

Each instance of MDPSP can be represented 
as a bipartite {\em hybridization graph} $G=((\bigcup_{i=1}^n P_i) \cup X, E)$,
with the left side containing all primers  
in the given pools and the right side containing the array probes, i.e., $X$. 
There is an edge between primer $p$ and probe $x\in X$ iff 
$x \in Spec_X(p,E_{p})$.
As discussed in Section \ref{sec.formulations}, 
we need to distinguish between the hybridizations that involve fluorescently 
labeled nucleotides and those that do not. 
Thus, for every primer $p$, we let 
$N^+(p)=Spec_X(p)$ and $N^-(p)=Spec_X(p,E_p) \setminus Spec_X(p)$. 
Similarly, for each probe $x\in X$, we 
let $N^+(x)=\{p |\ x \in N^+(p)\}$ and $N^-(x)=\{p |\ x \in N^-(p)\}$. 

We considered two versions of the Min-Greedy algorithm when run 
on the bipartite hybridization graph, depending on the side from which 
the minimum degree vertex is picked.  In the first version, 
referred to as MinPrimerGreedy,  we 
pick first a minimum degree node from the primers side, while in the second 
version, referred to as MinProbeGreedy, 
we pick first a minimum degree node from the probes side.
Thus, MinPrimerGreedy greedy picks at each step a minimum degree primer $p$ 
and pairs it with a minimum degree probe $x \in N^+(p)$. 
MinProbeGreedy greedy, selects at each step a minimum degree probe $x$ 
and pairs it with a minimum degree primer $p$ in $N^+(x)$. 
In both algorithms, all neighbors of $p$ and $x$ and 
their incident edges are removed from $G$. 
Also, at each step, the algorithms remove all vertices $u$, 
for which $N^+(u)=\emptyset$. 
These deletions ensure that
the primers $p$ selected at each step satisfy condition 
(\ref{4-color-strong}).
Both algorithms stop when the graph becomes empty. 

As described so far, the MinPrimerGreedy and MinProbeGreedy 
algorithms work when each pool 
contains only one primer and when the redundancy is 1. 
We extended the two variants to handle pools of size 
greater than 1 by simply removing from the graph all primers 
$p' \in P\setminus \{p\}$ when picking primer $p$ from pool $P$.
If the redundancy $r$ is greater than 1, then whenever we pick 
a primer $p$, we also pick it's $r$ probe neighbors from $N^+(p)$ 
with the smallest degrees (breaking ties arbitrarily).
The primer neighbors of all these $r$ probes will then be deleted 
from the graph. Moreover, the algorithm maintains the invariant that
$|N^+(p)| \geq r$ for every primer $p$ and $|N^+(x)| \geq 1$ for every probe $x$ 
by removing primers/probes for which the degree decreases below these bounds.
Figures \ref{fig.min-primer} and \ref{fig.min-probe}
give the pseudocode for the MinPrimerGreedy, 
respectively the MinProbeGreedy greedy algorithms. 
For the sake of clarity, they use two subroutines 
for removing a primer vertex, respectively a probe vertex, 
which are described in Figures \ref{fig.remove-primer} 
and \ref{fig.remove-probe}.

\begin{figure}[t]             
\begin{center}
{\footnotesize
\fbox{
\begin{minipage}{\textwidth}             
\begin{tabbing}
\hspace*{5mm}\=\hspace{5mm}\=\hspace{5mm}\=\hspace{5mm}\=\hspace{5mm}\=\hspace{5mm}\=  \kill
{\tt remove-primer} $(p)$ \\
\rule[3pt]{0.95\textwidth}{0.3pt} \\
{\tt Begin} \\
\>\>{\tt For all} $x \in N^+(p)$ {\tt do} \\
\>\>\> $N^+(x) \leftarrow N^+(x) \setminus \{p\}$ \\
\>\>\>  {\tt If} $|N^+(x)| = 0$ \\
\>\>\>  {\tt Then remove-probe} $(x)$ \\
\>\>\>  {\tt End If} \\ 
\>\>{\tt End For} \\
\>\>{\tt For all} $x \in N^-(p)$ {\tt do} \\
\>\>\> $N^-(x) \leftarrow N^-(x) \setminus \{p\}$ \\
\>\>{\tt End For} \\
\>\>Delete vertex $p$ from graph $G$ \\
{\tt End}
\end{tabbing}
\end{minipage}}
}
\caption{\label{fig.remove-primer} The remove-primer subroutine.}
\end{center}
\end{figure}

\begin{figure}[t]             
\begin{center}
{\footnotesize
\fbox{
\begin{minipage}{\textwidth}             
\begin{tabbing}
\hspace*{5mm}\=\hspace{5mm}\=\hspace{5mm}\=\hspace{5mm}\=\hspace{5mm}\=\hspace{5mm}\=  \kill
{\tt remove-probe} $(x)$ \\
\rule[3pt]{0.95\textwidth}{0.3pt} \\
{\tt Begin} \\
\>\>{\tt For all} $p \in N^+(x)$ {\tt do} \\
\>\>\> $N^+(p) \leftarrow N^+(p) \setminus \{x\}$ \\
\>\>\>  {\tt If} $|N^+(p)| < r$ \\
\>\>\>  {\tt Then} {\tt remove-primer} $(p)$ \\
\>\>\>  {\tt End If} \\ 
\>\>{\tt End For} \\
\>\>{\tt For all} $p \in N^-(x)$ {\tt do} \\
\>\>\> $N^-(p) \leftarrow N^-(p) \setminus \{x\}$ \\
\>\>{\tt End For} \\
\>\>Delete vertex $x$ from graph $G$ \\
{\tt End} \\
\end{tabbing}
\end{minipage}}
}
\caption{\label{fig.remove-probe} The remove-probe subroutine.}
\end{center}
\end{figure}

\begin{figure}[t]             
{\footnotesize
\fbox{
\begin{minipage}{\textwidth}             
\begin{tabbing}
\hspace*{5mm}\=\hspace{5mm}\=\hspace{5mm}\=\hspace{5mm}\=\hspace{5mm}\=\hspace{5mm}\=  \kill
{\tt Input:} Pools $\cP=\{P_1,\dots,P_n\}$, 
extension sets $E_p$, $p\in \cup_{i=1}^n P_i$, probe set $X$, and redundancy $r$\\
{\tt Output:} Strongly $r$-decodable subset of pools $\cP' \subseteq \cP$ 
and set $R$ of representative primers for the pools in \cP' \\
\rule[3pt]{0.95\textwidth}{0.3pt} \\
Construct hybridization graph $G$ \\
$\cP' \leftarrow \emptyset$ \\
$R \leftarrow \emptyset$ \\
{\tt While} $G$ is not empty {\tt do} \\
\>\> Find a minimum degree primer $p$, and let $P$ be the pool of $p$ \\
\>\> $\cP' \leftarrow \cP' \cup \{P\}$ \\
\>\>  $R \leftarrow R \cup \{p\}$  \\
\>\>  {\tt For each} $(p') \in P\setminus \{p\}$ {\tt do} \\
  \>\>\>  {\tt remove-primer}$(p')$ \\
\>\>  {\tt End For} \\
\>\> Let $|N^+(p)|=k$ and let $\{x_1,\ldots,x_k\}$ be the probes in $N^+(p)$, 
       indexed in increasing order of their degrees \\
\>\>  {\tt For each} $x \in \{x_1,\ldots,x_r\}$ {\tt do} \\
  \>\>\>  {\tt For each} $(p') \in N^+(x) \cup N^-(x)$ {\tt do} \\
  \>\>\>\>  {\tt remove-primer}$(p')$ \\
  \>\>\>  {\tt End For} \\
  \>\>\>  Delete vertex $x$ from $G$ \\ 
\>\>  {\tt End For} \\ 
\>\>  {\tt For each} $x \in \{x_{r+1},\ldots,x_k\} \cup N^-(p)$ {\tt do} \\
  \>\>\>  {\tt remove-probe}$(x)$ \\ 
\>\>  {\tt End For} \\
{\tt End While} \\
\end{tabbing}
\end{minipage}}
}
\caption{\label{fig.min-primer} MinPrimerGreedy greedy algorithm.}
\end{figure}

\begin{figure}[t]             
{\footnotesize
\fbox{
\begin{minipage}{\textwidth}             
\begin{tabbing}
\hspace*{5mm}\=\hspace{5mm}\=\hspace{5mm}\=\hspace{5mm}\=\hspace{5mm}\=\hspace{5mm}\=  \kill
{\tt Input:} Pools $\cP=\{P_1,\dots,P_n\}$, 
extension sets $E_p$, $p\in \cup_{i=1}^n P_i$, probe set $X$, and redundancy $r$\\
{\tt Output:} Strongly $r$-decodable subset of pools $\cP' \subseteq \cP$ 
and set $R$ of representative primers for the pools in \cP' \\
\rule[3pt]{0.95\textwidth}{0.3pt} \\
Construct hybridization graph $G$ \\
$\cP' \leftarrow \emptyset$ \\
$R \leftarrow \emptyset$ \\
{\tt While} $G$ is not empty {\tt do} \\
\>\> Find a minimum degree probe $x$ \\
\>\> Find a minimum degree primer $p$ in $N^+(x)$, and let $P$ be the pool of $p$ \\
\>\> $\cP' \leftarrow \cP' \cup \{P\}$ \\
\>\>  $R \leftarrow R \cup \{p\}$ \\ 
\>\>  {\tt For each} $p' \in P\setminus\{p\}$ {\tt do} \\
  \>\>\>  {\tt remove-primer}$(p')$ \\
\>\>  {\tt End For} \\
\>\> Let $|N^+(p)|=k$ and let $\{x_1,\ldots,x_k\}$ be the probes in $N^+(p)$, 
       indexed in increasing order of their degrees \\
\>\>  {\tt For each} $x \in \{x_1,\ldots,x_r\}$ {\tt do} \\
  \>\>\>  {\tt For each} $p' \in N^+(x) \cup N^-(x)$ {\tt do} \\
  \>\>\>\>  {\tt remove-primer}$(p')$ \\
  \>\>\>  {\tt End For} \\
  \>\>\>  Delete vertex $x$ from $G$ \\ 
\>\>  {\tt End For} \\ 
\>\>  {\tt For each} $x \in \{x_{r+1},\ldots,x_k\} \cup N^-(p)$ {\tt do} \\
  \>\>\>  {\tt remove-probe}$(x)$ \\ 
\>\>  {\tt End For} \\
{\tt End While} \\
\end{tabbing}
\end{minipage}}
}
\caption{\label{fig.min-probe} MinProbeGreedy greedy algorithm.}
\end{figure}

Algorithms MinPrimerGreedy and MinProbeGreedy can be implemented 
efficiently using a Fibonacci heap for maintaining the degrees 
of primers, respectively of probes. 
Let $N$ be the total number of primers in the $n$ pools, 
$m$ be the number of probes in $X$, and 
$k$ be the size of the $r$-decodable set returned by the 
algorithm.
Since each primer has bounded degree, the sorting of probe degrees 
requires $O(k)$ total time.  The total number of edges in the 
hybridization graph is $O(N+m)$. 
By using a Fibonacci heap, finding a minimum degree primer (probe)
can be done in $O(\log{N})$ (respectively $O(\log{m})$) and each 
primer degree update can be done in amortized $O(1)$ time.
Thus, the total runtime for MinPrimerGreedy algorithm is 
$O(k\log{N}+N+m)$, and 
the total runtime for  MinProbeGreedy algorithm is          
$O(k\log{m}+N+m)$.

\section{Experimental Results} 
\label{sec.results}

We considered 
two types of data sets:
\begin{itemize}
\item 
Randomly generated datasets containing between 1,000 to 200,000 
pools with 1 or 2 primers of length between 10 and 30. 
\item
Two-primer pools representing over 9 million reference SNPs 
in human chromosomes 1-22, X, and Y
extracted from the NCBI dbSNP database build 125.
We disregarded reference SNPs for which available 
flanking sequence was insufficient for 
determining two non-degenerate primers of desired length 
(due, e.g., to the presence of degenerate bases near the SNP locus).
\end{itemize}

We used two types of array probe sets. First, we used 
probe sets containing all $k$-mers, for $k$ between 8 and 10.
All $k$-mer arrays are well studied in the context of sequencing by 
hybridization. 
However, a major drawback of all $k$-mer arrays is that the 
$k$-mers have a wide range of melting temperatures, making 
it difficult to ensure reliable hybridization results.  
For short oligonucleotides, 
a good approximation of the melting temperature is obtained 
using the simple 2-4 rule of Wallace \cite{Wallace}, 
according to which the melting temperature 
of a probe is approximately twice the number of A and T bases, plus 
four times the number of C and G bases.  
As in \cite{Karp00}, we define the {\em weight} of a DNA string to be 
the number of A and T bases plus twice the number of C and G bases.  
For a given integer $c$, a DNA string is called a $c$-token if 
it has a weight $c$ or more and all its proper suffixes have   
weight strictly less than $c$.  
Since the weight of a $c$-token 
is either $c$ or $c+1$, it follows that the 2-4 rule computed 
melting temperature of all $c$-tokens varies in a range of about 
$4^{\circ}$C. 
In our experiments we used probe sets 
consisting of all $c$-tokens, with $c$ varying between 11 and 13. 
The considered values of $k$ and $c$ were picked such 
that the resulting number of probes is representative of 
current array manufacturing technologies: there are roughly 65,000 
8-mers, 262,000 9-mers, 1 million 10-mers,  86,000 11-tokens, 
236,000 12-tokens, and 645,000 13-tokens -- the smaller probe sets 
can be spotted using current oligonucleotide printing robots, while 
the larger probe sets can be synthesized in situ using photolithographic 
techniques. 

\subsection{Results on Synthetic Datasets}
\label{sec.random}

In a first set of experiments on the randomly generated datasets 
we compared the three MDPSP algorithms 
on instances with primer length set to 20, which is the 
typical length used, e.g., in genotyping using 
tag arrays.  In these experiments the set of possible extensions 
was considered to be $\{$A,C,T,G$\}$ for all primers. Such a conservative choice 
gives an estimate of multiplexing rates achievable by SBE/SBH assays 
in more demanding genomic analyses such as microorganism identification by 
DNA barcoding \cite{ijbra05}, in which a primer (typically 
referred to as a {\em distinguisher} in this context) may be extended 
by any of the DNA bases in different microorganisms. 
The results of these experiments for all $k$-mer and all $c$-token 
probe sets are presented in Tables \ref{table.set1.k} and 
\ref{table.set1.c}, respectively.
The results show that using the flexibility of picking primers 
from either strand of the genomic sequence yields an improvement 
of up to 10\% in the number of $r$-decodable pools. 
The MinProbeGreedy algorithm typically produces better 
results compared to the MinPrimerGreedy variant. 
On the other hand, neither Sequential Greedy nor MinProbeGreedy dominates the other 
algorithms for all range of instance parameters -- 
Sequential Greedy generally gives the better results for $k$-mer experiments with 
high redundancy values, while MinProbeGreedy generally gives better results for 
$k$-mer experiments with large number of pools and low redundancy 
and for $c$-token experiments.

In the second set of experiments we ran the three MDPSP algorithms on 
datasets with the same primer length of 20, pool size of 2, and 
with the number of possible extensions of each primer set to 4 
as in DNA-barcoding applications, and to 2 as in SNP genotyping.
The results for all $k$-mer and all $c$-token 
probe sets are given in Tables \ref{table.set2.k} and 
\ref{table.set2.c}.  The relative performance of the algorithms 
is similar to that observed in the first set of experiments.  
As expected, taking into account the reduced number of possible extensions 
increases the size of computed decodable pool subsets, often by more than 5\%.

\begin{figure}[ht]
{
\centerline{\psfig{figure=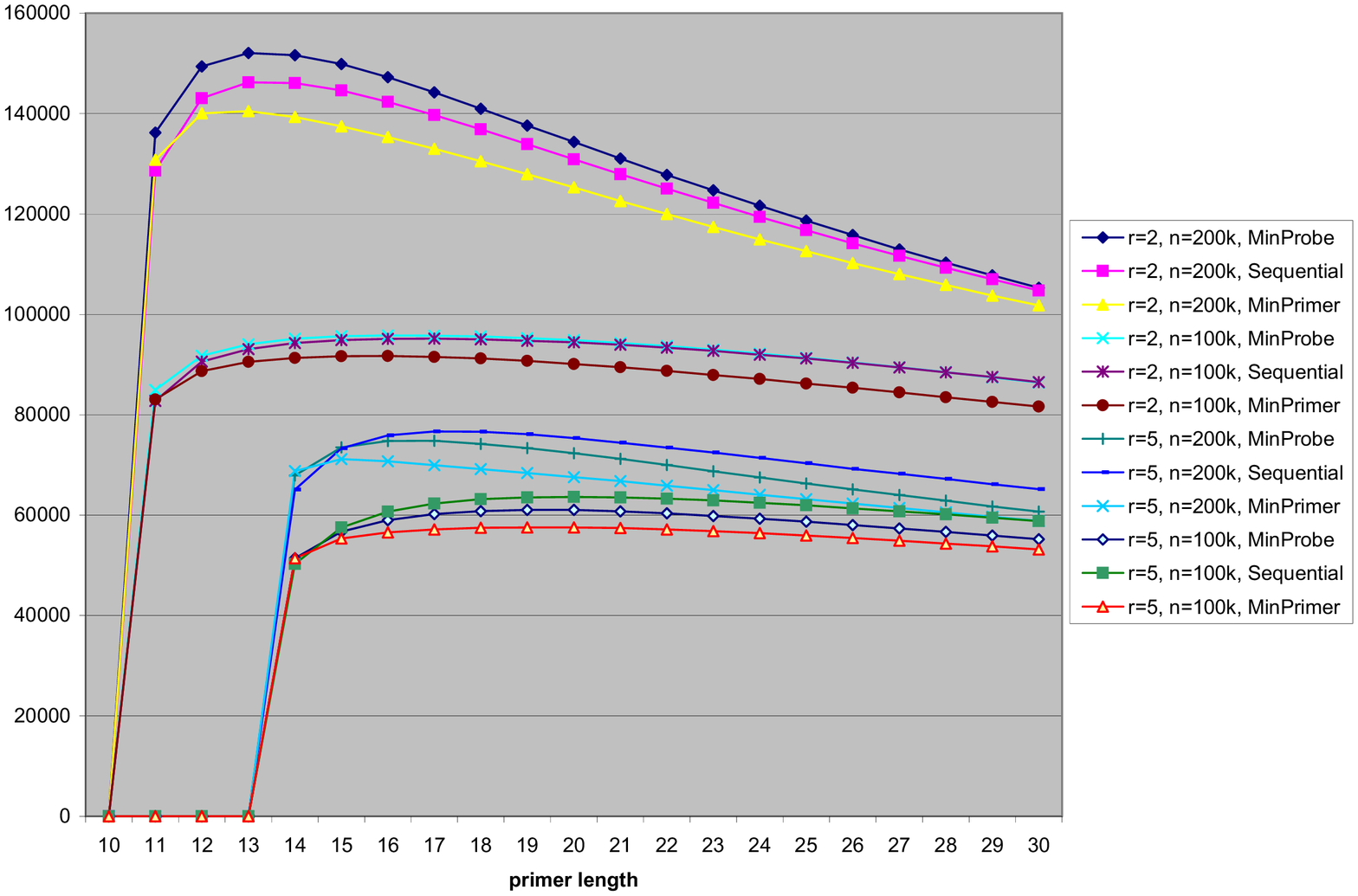,width=6in}}
\vspace{-0.5in}
\centerline{(a)}
\vspace{-0.3in}
\centerline{\psfig{figure=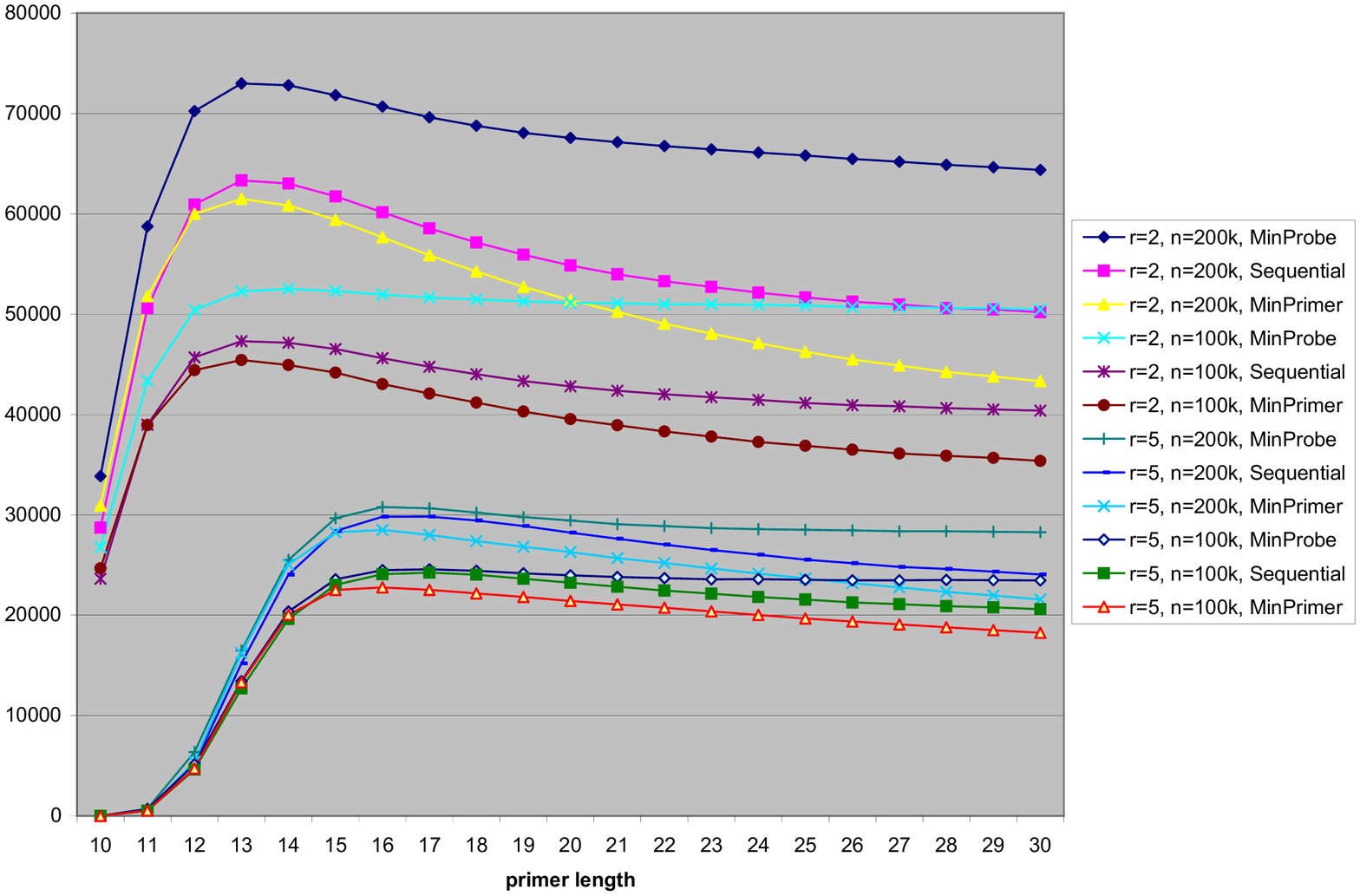,width=6in}}
\vspace{-0.5in}
\centerline{(b)}
\vspace{-0.1in}
\caption{\label{fig.length}
Size of the strongly $r$-decodable pool subset computed by the three        
MDPSP algorithms as a function of primer length, for 
pools with 2 primers, 2 possible extensions per primer, 
and array probes consisting of all $4^{10}$ 10-mers (a), respectively 
all 645,376 13-tokens (b) (averages over 10 test cases).}
}
\end{figure}

In the third set of experiments we explored the degree of freedom 
given by the primer length.  For any fixed array probe set and redundancy 
requirement, we need a minimum primer length to be able to 
satisfy constraints (\ref{4-color-strong}).  Increasing the primer length 
beyond this minimum primer length is often beneficial, as it increases 
the number of array probes that hybridize with the primer.  However, if 
primer length increases too much, an increasing number of these 
probes become non-specific, and the multiplexing rate starts to decline.
Figure \ref{fig.length} gives the 
tradeoff between primer length and the size of the 
strongly $r$-decodable pool subsets computed by the three MDPSP 
algorithms for pools with 2 primers, 2 possible extensions per primer and
all 10-mers, respectively all 13-tokens, as array probes. We notice that 
the optimal primer length increases with the redundancy parameter.

\begin{table}[t]
{\footnotesize
\begin{center}
\caption{\label{table.set1.k}
Size of the strongly $r$-decodable pool subset computed by the three 
MDPSP algorithms 
for primer length 20 and set of possible extensions $\{$A,C,T,G$\}$, 
with redundancy $r\in\{1, 2, 5\}$ and all $k$-mer 
probe sets for $k\in\{8, 9, 10\}$ (averages over 10 test cases).}
{\scriptsize
\begin{tabular}{| c | c | c | c c | c c | c c |}
\hline
r & \#  &  Algorithm &                 
\multicolumn{2}{c|}{k=8} &
\multicolumn{2}{c|}{k=9} &   
\multicolumn{2}{c|}{k=10}    
\\
 & pools &   &       
1 primer & 2 primers & 
1 primer & 2 primers & 
1 primer & 2 primers \\ 
\hline
\hline
       &       &Sequential     &1000   &1000   &1000   &1000   &1000   &1000
\\
       &1000   &MinPrimer      &1000   &1000   &1000   &1000   &1000   &1000
\\
       &       &MinProbe       &1000   &1000   &1000   &1000   &1000   &1000
\\
\cline{2-9}
       &       &Sequential     &2000   &2000   &2000   &2000   &2000   &2000
\\
       &2000   &MinPrimer      &2000   &2000   &2000   &2000   &2000   &2000
\\
       &       &MinProbe       &2000   &2000   &2000   &2000   &2000   &2000
\\
\cline{2-9}
       &       &Sequential     &7740   &8574   &9991   &10000  &10000  &10000
\\
       &10000  &MinPrimer      &7714   &8319   &9991   &9999   &10000  &10000
\\
1       &       &MinProbe       &7768   &8803   &9991   &10000  &10000  &10000
\\
\cline{2-9}
       &       &Sequential     &9967   &11071  &19436  &19948  &19999  &20000
\\
       &20000  &MinPrimer      &9889   &10999  &19447  &19745  &19999  &20000
\\
       &       &MinProbe       &9886   &11107  &19458  &19989  &19999  &20000
\\
\cline{2-9}
       &       &Sequential     &12486  &12656  &43279  &47688  &93632  &98630
\\
       &100000 &MinPrimer      &13864  &15324  &42980  &48021  &93642  &96712
\\
       &       &MinProbe       &13993  &15672  &43273  &48418  &93837  &99601
\\
\cline{2-9}
       &       &Sequential     &12635  &12658  &49062  &51646  &140820 &157908
\\
       &200000 &MinPrimer      &15476  &17010  &50347  &56017  &139787 &154028
\\
       &       &MinProbe       &15822  &17630  &50459  &56676  &141614 &160532
\\
\hline
\hline
       &       &Sequential     &1000   &1000   &1000   &1000   &1000   &1000
\\
       &1000   &MinPrimer      &1000   &1000   &1000   &1000   &1000   &1000
\\
       &       &MinProbe       &1000   &1000   &1000   &1000   &1000   &1000
\\
\cline{2-9}
       &       &Sequential     &1997   &2000   &2000   &2000   &2000   &2000
\\
       &2000   &MinPrimer      &1997   &2000   &2000   &2000   &2000   &2000
\\
       &       &MinProbe       &1997   &2000   &2000   &2000   &2000   &2000
\\
\cline{2-9}
       &       &Sequential     &6210   &6901   &9934   &9999   &10000  &10000
\\
       &10000  &MinPrimer      &6002   &6463   &9932   &9977   &10000  &10000
\\
2       &       &MinProbe       &6174   &6890   &9938   &9998   &10000  &10000
\\
\cline{2-9}
       &       &Sequential     &7463   &8192   &17948  &19274  &19992  &20000
\\
       &20000  &MinPrimer      &7052   &7662   &17812  &18455  &19992  &20000
\\
       &       &MinProbe       &7435   &8068   &18004  &19288  &19993  &20000
\\
\cline{2-9}
       &       &Sequential     &9254   &9644   &31845  &34855  &82315  &90627
\\
       &100000 &MinPrimer      &8917   &9605   &30043  &32700  &81056  &85852
\\
       &       &MinProbe       &9404   &10273  &31805  &34481  &82522  &90935
\\
\cline{2-9}
       &       &Sequential     &9674   &9953   &35514  &37891  &109450 &122470
\\
       &200000 &MinPrimer      &9658   &10333  &33479  &36247  &104891 &114624
\\
       &       &MinProbe       &10326  &11246  &35228  &38498  &109252 &122986
\\
\hline
\hline
       &       &Sequential     &995    &1000   &1000   &1000   &1000   &1000
\\
       &1000   &MinPrimer      &995    &999    &1000   &1000   &1000   &1000
\\
       &       &MinProbe       &995    &1000   &1000   &1000   &1000   &1000
\\
\cline{2-9}
       &       &Sequential     &1872   &1973   &1998   &2000   &2000   &2000
\\
       &2000   &MinPrimer      &1860   &1898   &1998   &2000   &2000   &2000
\\
       &       &MinProbe       &1866   &1946   &1998   &2000   &2000   &2000
\\
\cline{2-9}
       &       &Sequential     &3745   &4161   &8674   &9483   &9972   &10000
\\
       &10000  &MinPrimer      &3376   &3635   &8484   &8881   &9969   &9998
\\
5       &       &MinProbe       &3480   &3845   &8564   &9233   &9970   &10000
\\
\cline{2-9}
       &       &Sequential     &4289   &4705   &12204  &13750  &19498  &19967
\\
       &20000  &MinPrimer      &3748   &4029   &11393  &12360  &19435  &19804
\\
       &       &MinProbe       &3943   &4286   &11680  &12960  &19468  &19931
\\
\cline{2-9}
       &       &Sequential     &5241   &5520   &17920  &19612  &52078  &59021
\\
       &100000 &MinPrimer      &4450   &4726   &15580  &16781  &47922  &52711
\\
       &       &MinProbe       &4818   &5171   &16521  &17990  &49329  &55573
\\
\cline{2-9}
       &       &Sequential     &5534   &5775   &19767  &21251  &62791  &70334
\\
       &200000 &MinPrimer      &4724   &4990   &16959  &18116  &56160  &61406
\\
       &       &MinProbe       &5177   &5531   &18175  &19757  &58565  &65344
\\
\hline
\end{tabular}
}
\end{center}
}
\end{table}

\begin{table}[t]
{\footnotesize
\begin{center}
\caption{\label{table.set1.c}
Size of the strongly $r$-decodable pool subset computed by the three               
MDPSP algorithms 
for primer length 20 and set of possible extensions $\{$A,C,T,G$\}$, 
with redundancy $r\in\{1, 2, 5\}$ and all $c$-token probe sets for 
$c\in\{11, 12, 13\}$ 
(averages over 10 test cases).}
\scriptsize
\begin{tabular}{| c | c | c | c c | c c | c c |}
\hline
r & \#  &  Algorithm &                 
\multicolumn{2}{c|}{c=11} &
\multicolumn{2}{c|}{c=12} &   
\multicolumn{2}{c|}{c=13}    
\\
 & pools &   &       
1 primer & 2 primers & 
1 primer & 2 primers & 
1 primer & 2 primers \\ 
\hline
\hline
       &       &Sequential     &991    &1000   &999    &1000   &1000   &1000
\\
       &1000   &MinPrimer      &992    &999    &999    &1000   &1000   &1000
\\
       &       &MinProbe       &993    &1000   &999    &1000   &1000   &1000
\\
\cline{2-9}
       &       &Sequential     &1881   &1982   &1986   &2000   &1999   &2000
\\
       &2000   &MinPrimer      &1890   &1959   &1987   &1998   &1999   &2000
\\
       &       &MinProbe       &1906   &1994   &1988   &2000   &1999   &2000
\\
\cline{2-9}
       &       &Sequential     &5745   &6993   &8006   &9218   &9420   &9927
\\
       &10000  &MinPrimer      &5556   &6401   &8005   &8782   &9472   &9801
\\
1       &       &MinProbe       &6385   &7972   &8436   &9688   &9550   &9980
\\
\cline{2-9}
       &       &Sequential     &7968   &9733   &12458  &15191  &16656  &18931
\\
       &20000  &MinPrimer      &7490   &8798   &12242  &14080  &16673  &18204
\\
       &       &MinProbe       &9190   &11548  &13684  &17094  &17430  &19613
\\
\cline{2-9}
       &       &Sequential     &13708  &16042  &26407  &32202  &45064  &56064
\\
       &100000 &MinPrimer      &12564  &14736  &24482  &29336  &42824  &51540
\\
       &       &MinProbe       &16820  &20277  &31414  &39202  &51448  &65877
\\
\cline{2-9}
       &       &Sequential     &16241  &18516  &33278  &39552  &61351  &76037
\\
       &200000 &MinPrimer      &14967  &17278  &30762  &36618  &57530  &70048
\\
       &       &MinProbe       &20574  &24329  &40580  &49300  &72230  &91488
\\
\hline
\hline
       &       &Sequential     &965    &998    &997    &1000   &1000   &1000
\\
       &1000   &MinPrimer      &965    &986    &997    &999    &1000   &1000
\\
       &       &MinProbe       &972    &998    &997    &1000   &1000   &1000
\\
\cline{2-9}
       &       &Sequential     &1711   &1905   &1940   &1995   &1995   &2000
\\
       &2000   &MinPrimer      &1697   &1815   &1942   &1981   &1995   &2000
\\
       &       &MinProbe       &1766   &1948   &1951   &1997   &1996   &2000
\\
\cline{2-9}
       &       &Sequential     &4216   &5107   &6578   &7891   &8616   &9611
\\
       &10000  &MinPrimer      &3926   &4571   &6344   &7252   &8572   &9214
\\
2       &       &MinProbe       &4876   &6059   &7138   &8610   &8896   &9783
\\
\cline{2-9}
       &       &Sequential     &5482   &6589   &9450   &11615  &14060  &16839
\\
       &20000  &MinPrimer      &5024   &5901   &8919   &10551  &13699  &15613
\\
       &       &MinProbe       &6635   &8151   &10796  &13540  &15152  &17980
\\
\cline{2-9}
       &       &Sequential     &8587   &9839   &17469  &20811  &32223  &39839
\\
       &100000 &MinPrimer      &7897   &9071   &16133  &19192  &30138  &36595
\\
       &       &MinProbe       &10990  &12695  &21738  &26341  &38246  &48131
\\
\cline{2-9}
       &       &Sequential     &9899   &11114  &21192  &24696  &41783  &50811
\\
       &200000 &MinPrimer      &9149   &10418  &19730  &23155  &39125  &47357
\\
       &       &MinProbe       &12782  &14541  &26957  &31714  &51198  &63112
\\
\hline
\hline
       &       &Sequential     &787    &906    &947    &992    &992    &1000
\\
       &1000   &MinPrimer      &767    &837    &941    &971    &992    &999
\\
       &       &MinProbe       &794    &905    &947    &990    &992    &1000
\\
\cline{2-9}
       &       &Sequential     &1187   &1433   &1646   &1870   &1914   &1991
\\
       &2000   &MinPrimer      &1112   &1284   &1600   &1753   &1903   &1960
\\
       &       &MinProbe       &1204   &1437   &1652   &1856   &1914   &1986
\\
\cline{2-9}
       &       &Sequential     &2262   &2713   &4046   &4988   &6284   &7662
\\
       &10000  &MinPrimer      &2067   &2467   &3732   &4495   &5939   &6976
\\
5       &       &MinProbe       &2363   &2875   &4154   &5118   &6324   &7651
\\
\cline{2-9}
       &       &Sequential     &2779   &3279   &5347   &6540   &9139   &11399
\\
       &20000  &MinPrimer      &2553   &2998   &4908   &5956   &8504   &10308
\\
       &       &MinProbe       &2957   &3562   &5520   &6808   &9222   &11530
\\
\cline{2-9}
       &       &Sequential     &4020   &4536   &8753   &10211  &17580  &21359
\\
       &100000 &MinPrimer      &3738   &4250   &8122   &9494   &16252  &19645
\\
       &       &MinProbe       &4509   &5208   &9284   &11078  &18048  &22119
\\
\cline{2-9}
       &       &Sequential     &4538   &5035   &10286  &11738  &21762  &25859
\\
       &200000 &MinPrimer      &4264   &4749   &9609   &11054  &20226  &24058
\\
       &       &MinProbe       &5221   &5926   &11149  &12986  &22602  &27186
\\
\hline
\end{tabular}
\end{center}
}
\end{table}

\begin{table}[t]
{\footnotesize
\begin{center}
\caption{\label{table.set2.k}
Size of the strongly $r$-decodable pool subset computed by the three               
MDPSP algorithms 
for primer length 20 and 2 primers per pool, with number 
of possible extensions $|E_p|\in\{2,4\}$, 
redundancy $r\in\{1, 2, 5\}$ and all $k$-mer probe sets for 
$k\in \{8, 9, 10\}$ (averages over 10 test cases).}
\scriptsize
\begin{tabular}{| c | c | c | c c | c c | c c |}
\hline
r & \#  &  Algorithm &                 
\multicolumn{2}{c|}{k=8} &
\multicolumn{2}{c|}{k=9} &   
\multicolumn{2}{c|}{k=10}    
\\
 & SNPs &   &       
$|E_p|=4$ & $|E_p|=2$ & 
$|E_p|=4$ & $|E_p|=2$ & 
$|E_p|=4$ & $|E_p|=2$ \\ 
\hline
\hline
       &       &Sequential     &1000   &1000   &1000   &1000   &1000   &1000
\\
       &1000   &MinPrimer      &1000   &1000   &1000   &1000   &1000   &1000
\\
       &       &MinProbe       &1000   &1000   &1000   &1000   &1000   &1000
\\
\cline{2-9}
       &       &Sequential     &2000   &2000   &2000   &2000   &2000   &2000
\\
       &2000   &MinPrimer      &2000   &2000   &2000   &2000   &2000   &2000
\\
       &       &MinProbe       &2000   &2000   &2000   &2000   &2000   &2000
\\
\cline{2-9}
       &       &Sequential     &8574   &8950   &10000  &10000  &10000  &10000
\\
       &10000  &MinPrimer      &8319   &8752   &9999   &10000  &10000  &10000
\\
1       &       &MinProbe       &8803   &9358   &10000  &10000  &10000  &10000
\\
\cline{2-9}
       &       &Sequential     &11071  &11673  &19948  &19981  &20000  &20000
\\
       &20000  &MinPrimer      &10999  &11898  &19745  &19873  &20000  &20000
\\
       &       &MinProbe       &11107  &12051  &19989  &19998  &20000  &20000
\\
\cline{2-9}
       &       &Sequential     &12656  &13813  &47688  &50643  &98630  &99478
\\
       &100000 &MinPrimer      &15324  &16551  &48021  &52263  &96712  &98209
\\
       &       &MinProbe       &15672  &16800  &48418  &52712  &99601  &99885
\\
\cline{2-9}
       &       &Sequential     &12658  &13890  &51646  &55694  &157908 &166796
\\
       &200000 &MinPrimer      &17010  &18216  &56017  &60962  &154028 &164696
\\
       &       &MinProbe       &17630  &18783  &56676  &61488  &160532 &173910
\\
\hline
\hline
       &       &Sequential     &1000   &1000   &1000   &1000   &1000   &1000
\\
       &1000   &MinPrimer      &1000   &1000   &1000   &1000   &1000   &1000
\\
       &       &MinProbe       &1000   &1000   &1000   &1000   &1000   &1000
\\
\cline{2-9}
       &       &Sequential     &2000   &2000   &2000   &2000   &2000   &2000
\\
       &2000   &MinPrimer      &2000   &2000   &2000   &2000   &2000   &2000
\\
       &       &MinProbe       &2000   &2000   &2000   &2000   &2000   &2000
\\
\cline{2-9}
       &       &Sequential     &6901   &7325   &9999   &10000  &10000  &10000
\\
       &10000  &MinPrimer      &6463   &6977   &9977   &9993   &10000  &10000
\\
2       &       &MinProbe       &6890   &7443   &9998   &9999   &10000  &10000
\\
\cline{2-9}
       &       &Sequential     &8192   &8639   &19274  &19670  &20000  &20000
\\
       &20000  &MinPrimer      &7662   &8348   &18455  &18988  &20000  &20000
\\
       &       &MinProbe       &8068   &8808   &19288  &19661  &20000  &20000
\\
\cline{2-9}
       &       &Sequential     &9644   &10175  &34855  &36886  &90627  &94420
\\
       &100000 &MinPrimer      &9605   &10398  &32700  &35771  &85852  &90098
\\
       &       &MinProbe       &10273  &11093  &34481  &37743  &90935  &94868
\\
\cline{2-9}
       &       &Sequential     &9953   &10535  &37891  &40060  &122470 &130911
\\
       &200000 &MinPrimer      &10333  &11143  &36247  &39619  &114624 &125287
\\
       &       &MinProbe       &11246  &12068  &38498  &41857  &122986 &134342
\\
\hline
\hline
       &       &Sequential     &1000   &1000   &1000   &1000   &1000   &1000
\\
       &1000   &MinPrimer      &999    &1000   &1000   &1000   &1000   &1000
\\
       &       &MinProbe       &1000   &1000   &1000   &1000   &1000   &1000
\\
\cline{2-9}
       &       &Sequential     &1973   &1989   &2000   &2000   &2000   &2000
\\
       &2000   &MinPrimer      &1898   &1933   &2000   &2000   &2000   &2000
\\
       &       &MinProbe       &1946   &1975   &2000   &2000   &2000   &2000
\\
\cline{2-9}
       &       &Sequential     &4161   &4405   &9483   &9722   &10000  &10000
\\
       &10000  &MinPrimer      &3635   &3970   &8881   &9211   &9998   &9999
\\
5       &       &MinProbe       &3845   &4204   &9233   &9546   &10000  &10000
\\
\cline{2-9}
       &       &Sequential     &4705   &4924   &13750  &14739  &19967  &19985
\\
       &20000  &MinPrimer      &4029   &4391   &12360  &13378  &19804  &19905
\\
       &       &MinProbe       &4286   &4690   &12960  &14110  &19931  &19973
\\
\cline{2-9}
       &       &Sequential     &5520   &5727   &19612  &20634  &59021  &63631
\\
       &100000 &MinPrimer      &4726   &5114   &16781  &18352  &52711  &57521
\\
       &       &MinProbe       &5171   &5581   &17990  &19741  &55573  &61043
\\
\cline{2-9}
       &       &Sequential     &5775   &5970   &21251  &22193  &70334  &75361
\\
       &200000 &MinPrimer      &4990   &5375   &18116  &19732  &61406  &67565
\\
       &       &MinProbe       &5531   &5939   &19757  &21555  &65344  &72313
\\
\hline
\end{tabular}
\end{center}
}

\end{table}
\begin{table}[t]
{\footnotesize
\begin{center}
\caption{\label{table.set2.c}
Size of the strongly $r$-decodable pool subset computed by the three               
MDPSP algorithms 
for primer length 20 and 2 primers per pool, with number              
of possible extensions $|E_p|\in\{2,4\}$,     
redundancy $r\in\{1, 2, 5\}$ and all $c$-token probe sets for 
$c\in\{11, 12, 13\}$ (averages over 10 test cases).}
\scriptsize
\begin{tabular}{| c | c | c | c c | c c | c c |}
\hline
r & \#  &  Algorithm &                 
\multicolumn{2}{c|}{c=11} &
\multicolumn{2}{c|}{c=12} &   
\multicolumn{2}{c|}{c=13}    
\\
 & SNPs &   &       
$|E_p|=4$ & $|E_p|=2$ & 
$|E_p|=4$ & $|E_p|=2$ & 
$|E_p|=4$ & $|E_p|=2$ \\ 
\hline
\hline
       &       &Sequential     &1000   &1000   &1000   &1000   &1000   &1000
\\
       &1000   &MinPrimer      &999    &999    &1000   &1000   &1000   &1000
\\
       &       &MinProbe       &1000   &1000   &1000   &1000   &1000   &1000
\\
\cline{2-9}
       &       &Sequential     &1982   &1990   &2000   &2000   &2000   &2000
\\
       &2000   &MinPrimer      &1959   &1968   &1998   &1998   &2000   &2000
\\
       &       &MinProbe       &1994   &1998   &2000   &2000   &2000   &2000
\\
\cline{2-9}
       &       &Sequential     &6993   &7324   &9218   &9412   &9927   &9953
\\
       &10000  &MinPrimer      &6401   &6776   &8782   &9034   &9801   &9866
\\
1       &       &MinProbe       &7972   &8280   &9688   &9782   &9980   &9990
\\
\cline{2-9}
       &       &Sequential     &9733   &10358  &15191  &15843  &18931  &19197
\\
       &20000  &MinPrimer      &8798   &9489   &14080  &14797  &18204  &18573
\\
       &       &MinProbe       &11548  &12187  &17094  &17599  &19613  &19746
\\
\cline{2-9}
       &       &Sequential     &16042  &17216  &32202  &34459  &56064  &59498
\\
       &100000 &MinPrimer      &14736  &15817  &29336  &31608  &51540  &55031
\\
       &       &MinProbe       &20277  &21599  &39202  &41665  &65877  &69188
\\
\cline{2-9}
       &       &Sequential     &18516  &19789  &39552  &42556  &76037  &81443
\\
       &200000 &MinPrimer      &17278  &18483  &36618  &39500  &70048  &75470
\\
       &       &MinProbe       &24329  &25757  &49300  &52534  &91488  &97154
\\
\hline
\hline
       &       &Sequential     &998    &998    &1000   &1000   &1000   &1000
\\
       &1000   &MinPrimer      &986    &990    &999    &1000   &1000   &1000
\\
       &       &MinProbe       &998    &999    &1000   &1000   &1000   &1000
\\
\cline{2-9}
       &       &Sequential     &1905   &1931   &1995   &1998   &2000   &2000
\\
       &2000   &MinPrimer      &1815   &1852   &1981   &1986   &2000   &2000
\\
       &       &MinProbe       &1948   &1962   &1997   &1999   &2000   &2000
\\
\cline{2-9}
       &       &Sequential     &5107   &5431   &7891   &8231   &9611   &9716
\\
       &10000  &MinPrimer      &4571   &4924   &7252   &7621   &9214   &9381
\\
2       &       &MinProbe       &6059   &6372   &8610   &8833   &9783   &9851
\\
\cline{2-9}
       &       &Sequential     &6589   &7036   &11615  &12312  &16839  &17409
\\
       &20000  &MinPrimer      &5901   &6388   &10551  &11255  &15613  &16231
\\
       &       &MinProbe       &8151   &8674   &13540  &14184  &17980  &18396
\\
\cline{2-9}
       &       &Sequential     &9839   &10552  &20811  &22486  &39839  &42814
\\
       &100000 &MinPrimer      &9071   &9819   &19192  &20864  &36595  &39542
\\
       &       &MinProbe       &12695  &13562  &26341  &28190  &48131  &51125
\\
\cline{2-9}
       &       &Sequential     &11114  &11894  &24696  &26659  &50811  &54858
\\
       &200000 &MinPrimer      &10418  &11212  &23155  &25122  &47357  &51390
\\
       &       &MinProbe       &14541  &15467  &31714  &34015  &63112  &67567
\\
\hline
\hline
       &       &Sequential     &906    &932    &992    &996    &1000   &1000
\\
       &1000   &MinPrimer      &837    &868    &971    &981    &999    &999
\\
       &       &MinProbe       &905    &928    &990    &994    &1000   &1000
\\
\cline{2-9}
       &       &Sequential     &1433   &1497   &1870   &1896   &1991   &1995
\\
       &2000   &MinPrimer      &1284   &1350   &1753   &1800   &1960   &1974
\\
       &       &MinProbe       &1437   &1511   &1856   &1885   &1986   &1990
\\
\cline{2-9}
       &       &Sequential     &2713   &2944   &4988   &5343   &7662   &8000
\\
       &10000  &MinPrimer      &2467   &2668   &4495   &4825   &6976   &7324
\\
5       &       &MinProbe       &2875   &3081   &5118   &5436   &7651   &7988
\\
\cline{2-9}
       &       &Sequential     &3279   &3552   &6540   &7040   &11399  &12143
\\
       &20000  &MinPrimer      &2998   &3273   &5956   &6424   &10308  &11007
\\
       &       &MinProbe       &3562   &3817   &6808   &7314   &11530  &12240
\\
\cline{2-9}
       &       &Sequential     &4536   &4912   &10211  &11140  &21359  &23232
\\
       &100000 &MinPrimer      &4250   &4610   &9494   &10352  &19645  &21421
\\
       &       &MinProbe       &5208   &5602   &11078  &11932  &22119  &23977
\\
\cline{2-9}
       &       &Sequential     &5035   &5443   &11738  &12809  &25859  &28234
\\
       &200000 &MinPrimer      &4749   &5128   &11054  &12022  &24058  &26297
\\
       &       &MinProbe       &5926   &6363   &12986  &13987  &27186  &29439
\\
\hline
\end{tabular}
\end{center}
}
\end{table}

\subsection{Results on dbSNP Data} 
\label{sec.dbSNP}

To stress-test our methods, we extracted a total of over 
9 million 2-primer pools corresponding to reference SNPs
in human chromosomes 1-22, X, and Y in the NCBI dbSNP 
database build 125. We constructed a dataset for each of the 
24 chromosomes by creating a 2-primer pool for each 
reference SNP for which dbSNP contains at least 
20 non-degenerate base pairs of flanking sequence 
on both sides (the number of reference SNPs and extracted pools 
for each chromosome are given in Table \ref{table.dbSNP}).
Since these large sets of pools must  be partitioned between multiple
SBE/SBH experiments, we used a simple MPPP algorithm which 
iteratively finds maximum $r$-decodable pool subsets using the sequential 
greedy algorithm.

\begin{figure}[!p]                 
\centerline{\psfig{figure=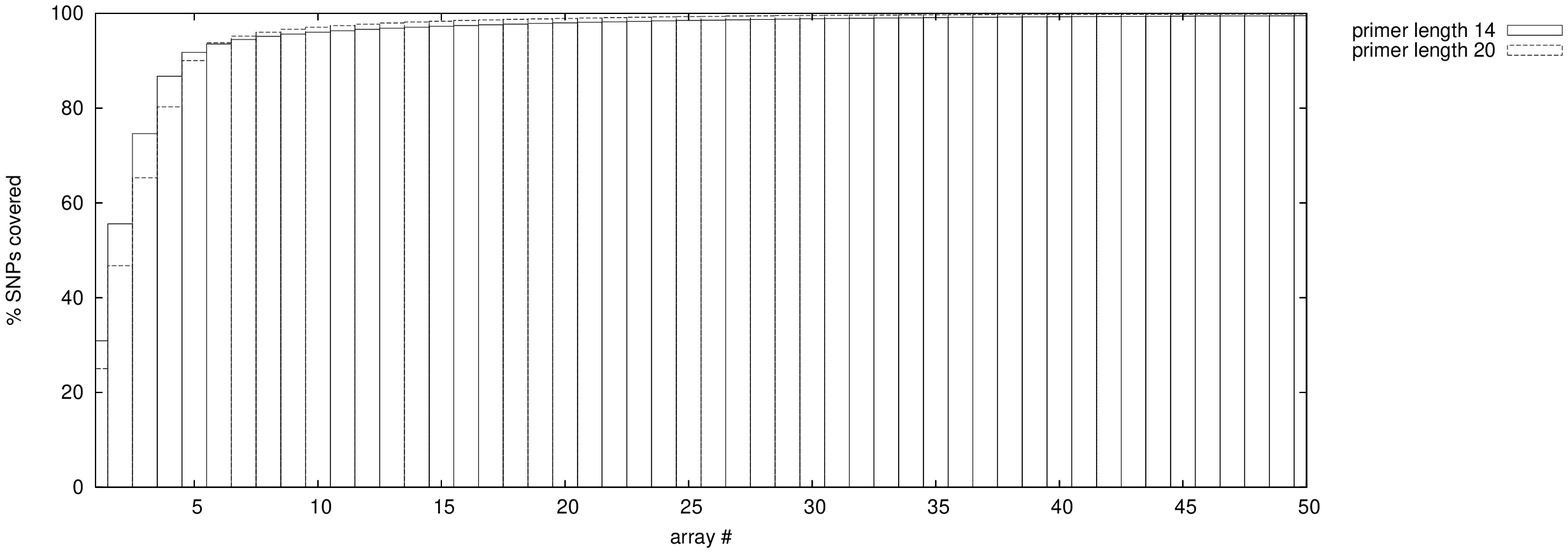,width=5.5in}}
\centerline{$r=1$}
\centerline{\psfig{figure=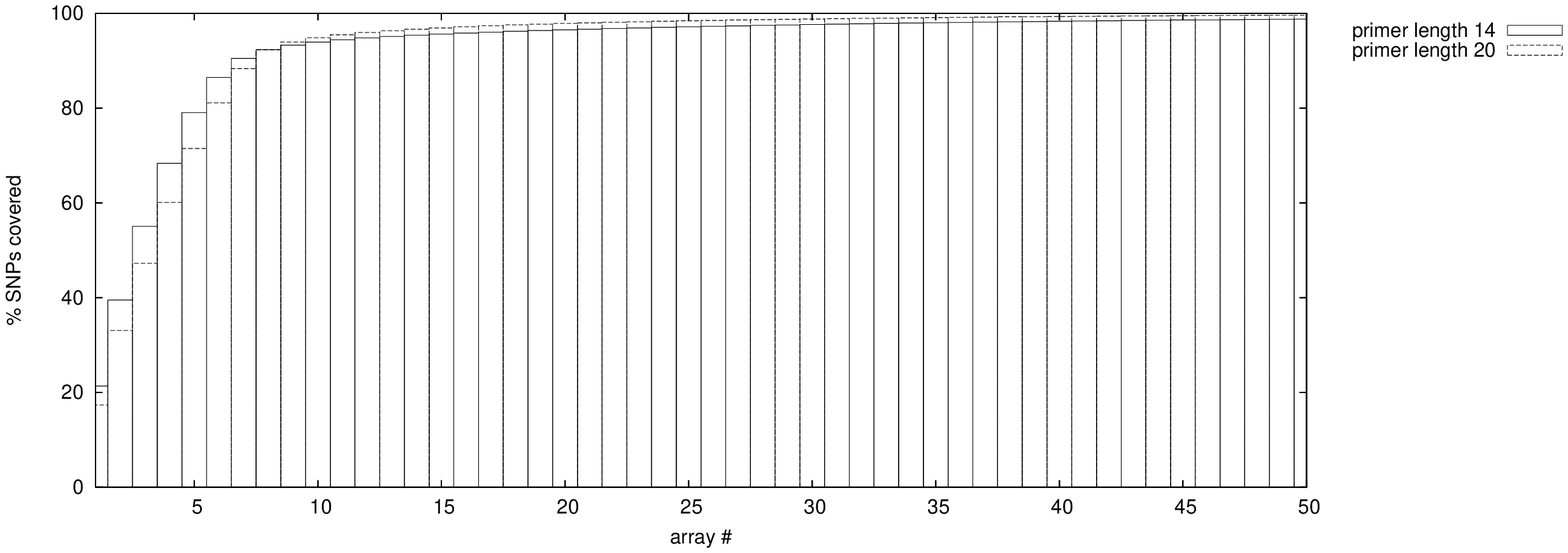,width=5.5in}}
\centerline{$r=2$}
\centerline{\psfig{figure=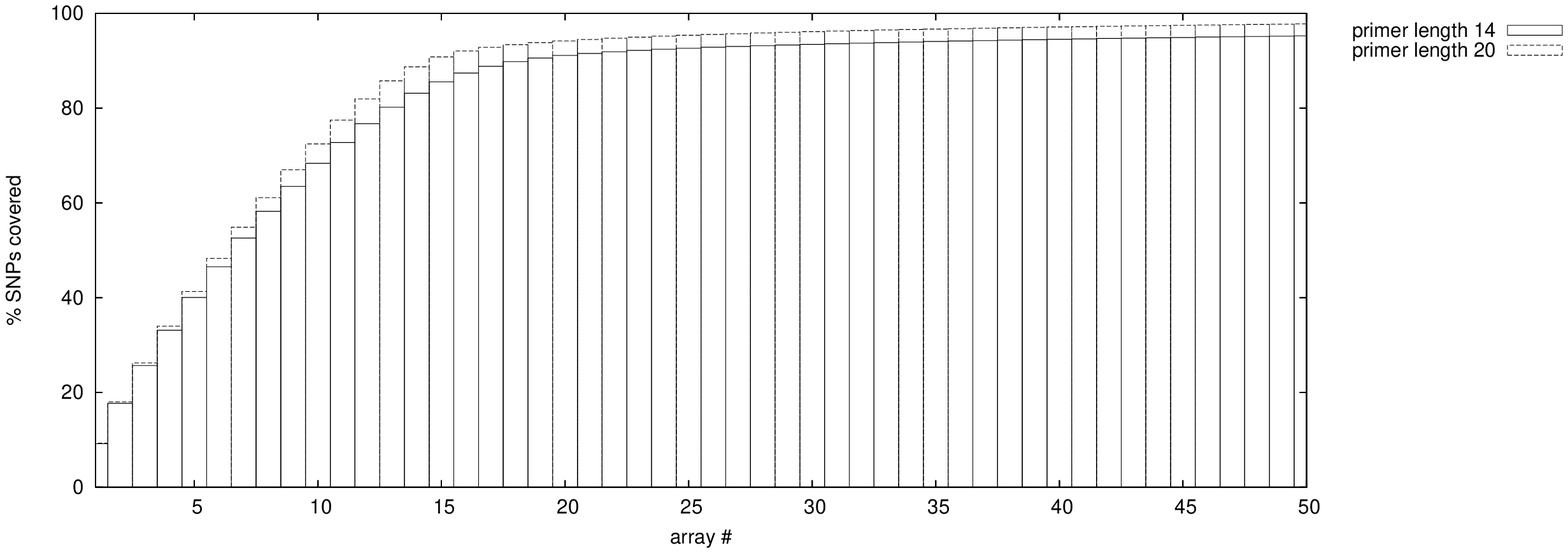,width=5.5in}}
\centerline{$r=5$}
\caption{\label{fig.chr1.k=10} Cumulative coverage rates for the first 50 10-mers arrays used to 
decode the SNPs in Chromosome 1 with primer length 14 or 20 and redundancy $r\in \{1, 2, 5\}$.
}
\end{figure}

\begin{figure}[!p]                 
\centerline{\psfig{figure=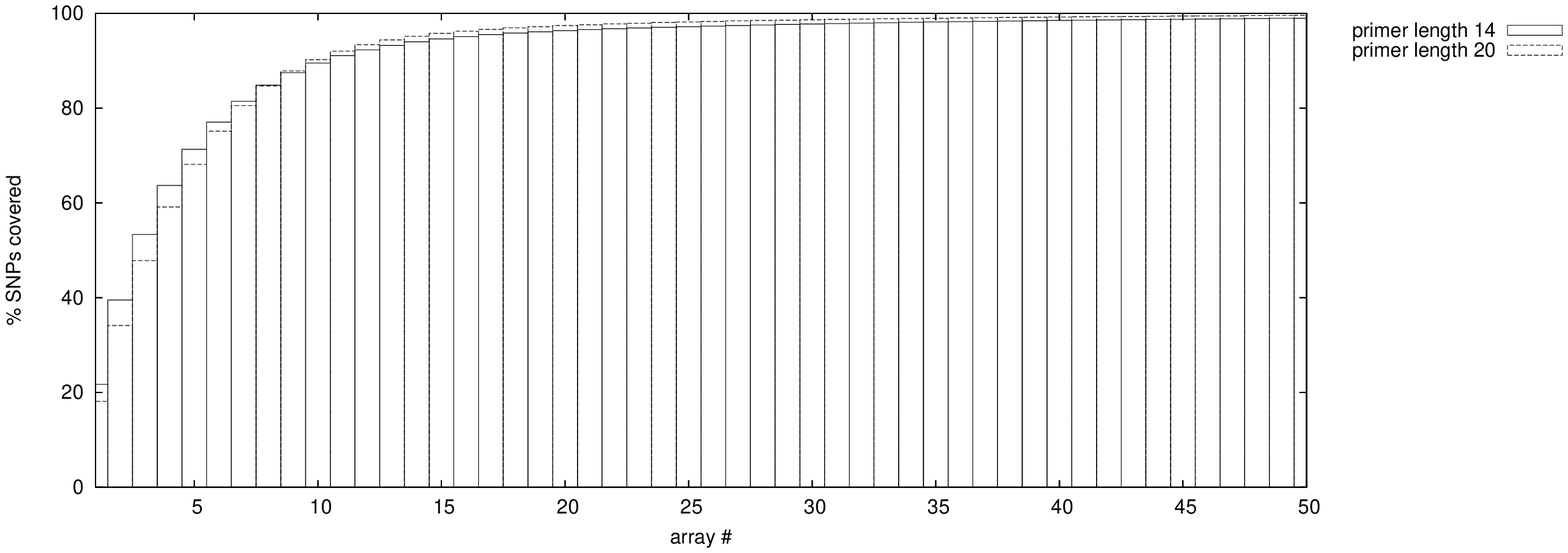,width=5.5in}}
\centerline{$r=1$}
\centerline{\psfig{figure=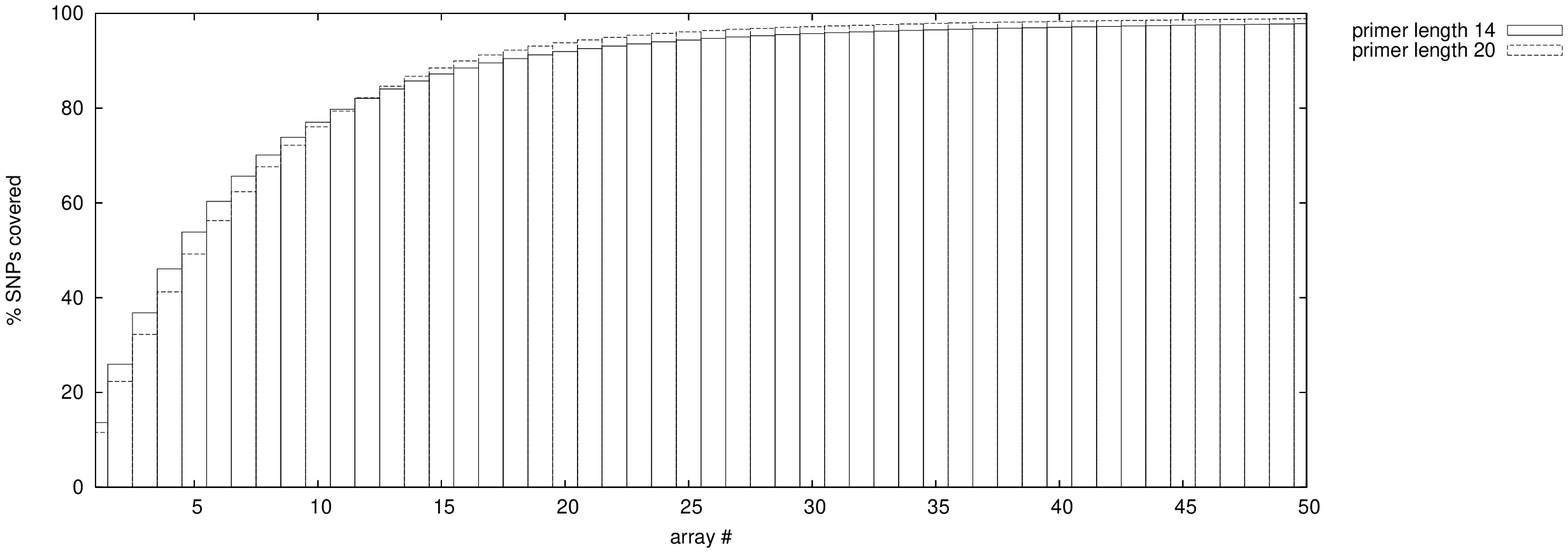,width=5.5in}}
\centerline{$r=2$}
\centerline{\psfig{figure=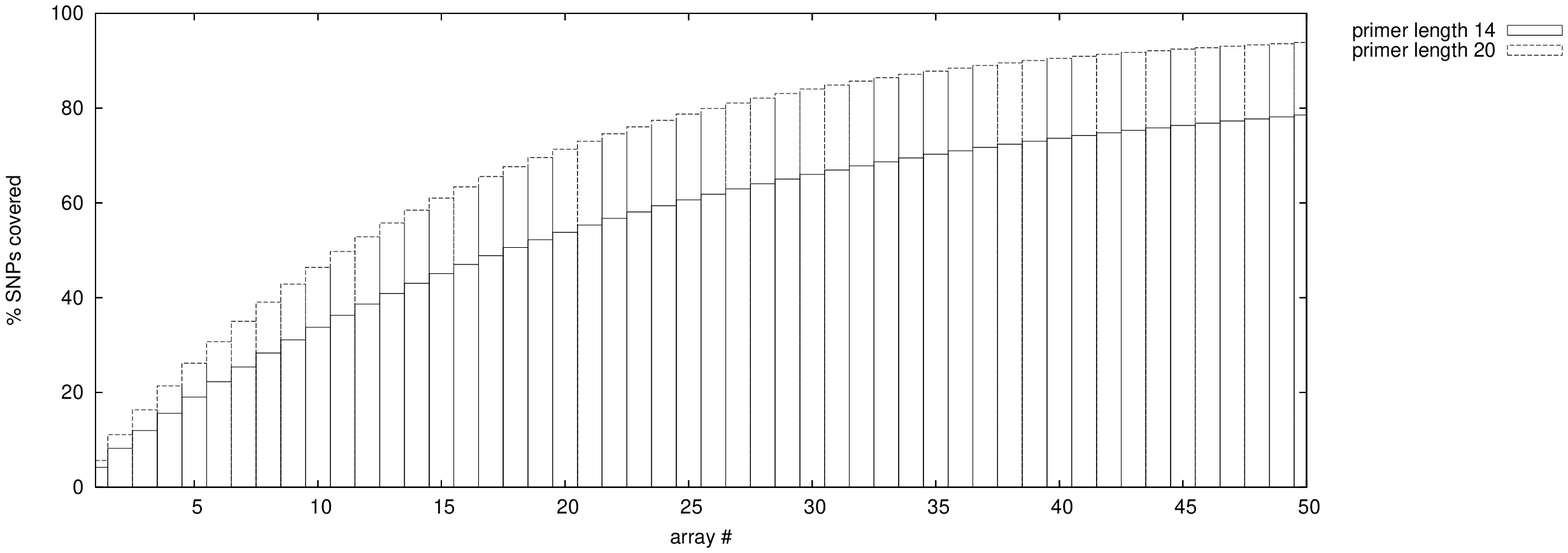,width=5.5in}}
\centerline{$r=5$}
\caption{\label{fig.chr1.c=13} Cumulative coverage rates for the first 50 13-tokens arrays used to 
decode the SNPs in Chromosome 1 with primer length 14 or 20 and redundancy $r\in \{1, 2, 5\}$.
}
\end{figure}

Figures \ref{fig.chr1.k=10} and \ref{fig.chr1.c=13} give the cumulative 
coverage percentage for the first 50 arrays of all 10-mers, respectively all 13-tokens, 
on the set of pools extracted from the human chromosome 1.  
In these experiments we used redundancy between 1 and 5, and primer length 
14 or 20.  While the MDPSP size in the first few iterations of our MPPP algorithm 
is comparable to those reported for randomly generated datasets 
in Section \ref{sec.random}, the number of SNPs assayed per array decreases 
constantly with array number -- as we need to assay more and more ``difficult'' SNPs. 
Somehow surprisingly, the results also suggest using primers of different lengths 
in different SBE/SBH experiments: while a primer length of 14 seems to be optimal 
for the first few arrays, longer primers improve the degree of multiplexing 
when only hard to differentiate SNPs remain, especially for high redundancy.  

Finally, in Table \ref{table.dbSNP} we give the number of arrays (containing either 
all 10-mers or all 13-tokens) required to cover $90\%$, respectively $95\%$ of 
the extracted reference SNPs, when using primers of length 20.  
In practical association studies a much lower SNP coverage (and hence much fewer 
arrays) would be required due to the high degree of linkage disequilibrium 
between the SNPs in the human population \cite{Patil01}.

\setlength\tabcolsep{4pt}

\begin{table}[t]
{\footnotesize
\begin{center}
\caption{\label{table.dbSNP}
Number of arrays needed to cover $90-95\%$ of the reference SNPs that 
have unambiguous primers of length 20.}
\begin{tabular}{| c | c | c | c c | c c | c c | c c | c c | c c |}
\hline
 Chr & \#  & \#  &
\multicolumn{6}{c|}{\# 10-mer arrays} &
\multicolumn{6}{c|}{\# 13-token arrays} \\
\cline{4-15}
 ID & Ref. & Extracted  &
\multicolumn{2}{c|}{r=1} &
\multicolumn{2}{c|}{r=2} &
\multicolumn{2}{c|}{r=5} & 
\multicolumn{2}{c|}{r=1} &
\multicolumn{2}{c|}{r=2} &
\multicolumn{2}{c|}{r=5}
\\
\cline{4-15}
 & SNPs & Pools & 
90\% & 95\% & 
90\% & 95\% & 
90\% & 95\% & 
90\% & 95\% & 
90\% & 95\% & 
90\% & 95\% \\
\hline
\hline
1	&786058	&736850	&5	&7	&8	&11	&15	&24	&10	&14	&17	&23	&39	&56	
\\
2	&758368	&704415	&5	&6	&7	&9	&14	&18	&9	&12	&14	&18	&32	&42	
\\
3	&647918	&587531	&5	&6	&7	&8	&13	&16	&8	&10	&12	&15	&26	&35	
\\
4	&690063	&646534	&5	&6	&7	&9	&14	&17	&8	&10	&12	&15	&26	&34	
\\
5	&590891	&550794	&5	&6	&6	&8	&12	&16	&7	&10	&12	&15	&26	&34	
\\
6	&791255	&742894	&10	&20	&14	&29	&30	&54	&15	&29	&23	&38	&49	&73	
\\
7	&666932	&629089	&6	&9	&8	&12	&16	&25	&10	&15	&16	&22	&36	&48	
\\
8	&488654	&456856	&4	&5	&5	&7	&10	&12	&7	&8	&10	&13	&22	&29	
\\
9	&465325	&441627	&4	&6	&6	&8	&11	&17	&7	&10	&11	&16	&26	&36	
\\
10	&512165	&480614	&4	&6	&6	&8	&11	&16	&8	&10	&12	&16	&27	&38	
\\
11	&505641	&476379	&4	&6	&6	&8	&11	&15	&8	&10	&12	&15	&26	&35	
\\
12	&474310	&443988	&4	&6	&6	&8	&11	&18	&7	&10	&11	&15	&25	&36	
\\
13	&371187	&347921	&3	&4	&5	&6	&9	&11	&5	&7	&8	&10	&16	&22	
\\
14	&292173	&271130	&3	&4	&4	&5	&7	&10	&5	&7	&8	&10	&16	&23	
\\
15	&277543	&258094	&3	&4	&4	&5	&7	&11	&5	&7	&8	&10	&17	&24	
\\
16	&306530	&288652	&4	&6	&5	&9	&9	&18	&7	&10	&11	&15	&25	&35	
\\
17	&269887	&249563	&3	&5	&4	&8	&9	&18	&7	&10	&11	&15	&25	&37	
\\
18	&268582	&250594	&3	&3	&4	&5	&7	&9	&4	&6	&6	&8	&14	&18	
\\
19	&212057	&199221	&4	&6	&5	&9	&11	&21	&8	&11	&12	&17	&29	&43	
\\
20	&292248	&262567	&3	&4	&4	&5	&7	&11	&6	&8	&9	&12	&20	&27	
\\
21	&148798	&138825	&2	&3	&3	&3	&5	&6	&3	&4	&5	&6	&10	&13	
\\
22	&175939	&164632	&3	&4	&3	&6	&6	&13	&6	&8	&9	&12	&21	&29	
\\
X	&380246	&362778	&4	&6	&6	&8	&10	&15	&6	&9	&9	&13	&19	&26	
\\
Y	&50725	&49372	&2	&2	&2	&2	&3	&3	&2	&2	&2	&3	&4	&5	
\\
\hline
\end{tabular}
\end{center}
}
\end{table}




\end{document}